\documentclass[review]{elsarticle}
\usepackage{graphicx}% Include figure files
\usepackage{dcolumn}% Align table columns on decimal point
\usepackage{bm}% bold mathf
\usepackage{braket}% bra-ket (Dirac) notation
\usepackage{graphicx}% Include figure files
\usepackage{epstopdf}
\usepackage{mwe}    % loads »blindtext« and »graphicx«
\usepackage{subfig}
\usepackage[T1]{fontenc}
\usepackage{caption}
\usepackage{amssymb}
\usepackage{amsmath}
\usepackage{booktabs}
\usepackage{tabularx}
\usepackage{multirow}

 \newcommand{\tensor}[1]{\bm{\mathsf{#1}}}
 
\begin{document}

%\preprint{PREPRINT}
\begin{frontmatter}
\title{Central Moments-based Cascaded Lattice Boltzmann Method for Thermal Convective Flows in Three-Dimensions
}
\author[label1]{Farzaneh  Hajabdollahi}
\ead{farzaneh.hajabdollahiouderji@ucdenver.edu}
\author[label1]{Kannan N. Premnath}
\ead{kannan.premnath@ucdenver.edu}
\address[label1]{Department of Mechanical Engineering, University of Colorado Denver, 1200 Larimer street, Colorado 80217 , U.S.A}

\date{\today}% It is always \today, today,

\begin{abstract}
Fluid motion driven by thermal effects, such as that due to buoyancy in differentially heated three-dimensional (3D) enclosures,  arise in several natural settings and engineering applications. It is represented by the solutions of the Navier-Stokes equations (NSE) in conjunction with the thermal energy transport equation represented as a convection-diffusion equation (CDE) for the temperature field. In this study, we develop new 3D lattice Boltzmann (LB) methods based on central moments and using multiple relaxation times for the three-dimensional, fifteen velocity (D3Q15) lattice, as well as it subset, i.e. the three-dimensional, seven velocity (D3Q7) lattice to solve the 3D CDE for the temperature field in a double distribution function framework. Their collision operators lead to a cascaded structure involving higher order terms resulting in improved stability. In this approach, the fluid motion is solved by another 3D cascaded LB model from prior work. Owing to the differences in the number of collision invariants to represent the dynamics of flow and the transport of the temperature field, the structure of the collision operator for the 3D cascaded LB formulation for the CDE is found to be markedly different from that for the NSE. The new 3D cascaded (LB) models for thermal convective flows are validated for natural convection of air driven thermally on two vertically opposite faces in a cubic cavity enclosure at different Rayleigh numbers against prior numerical benchmark solutions. Results show good quantitative agreement of the profiles of the flow and thermal fields, and the magnitudes of the peak convection velocities as well as the heat transfer rates given in terms of the Nusselt number.
\end{abstract}
\end{frontmatter}

\section{\label{app:Intro}Introduction}
 The thermal energy transport equation for convective flows, represented by means of a convection-diffusion equation (CDE) for the temperature field, can be classified as a combined hyperbolic and parabolic type partial differential equation (PDE). Solution of an such equation has received considerable attention for its key role in the study of many transport phenomena arising in various thermal science and engineering applications. In addition, the CDE-type models represent several important associated physical phenomena in fluid dynamics, such as the transport of the concentration of a chemical species as a passive scalar, and in the implicit capturing of interfaces in multiphase flows represented by phase field models. Whereas only for relatively simple geometries and boundary conditions, and under idealized physical situations exact analytical solutions of such equations are available, the development and applications of numerical methods play an essential role in investigations related to thermal convective flows, especially in three-dimensions (3D). Numerical techniques such as the finite difference, finite volume and finite element methods based on the direct discretization of the continuum PDEs such as the CDE have been investigated in the past. From a different perspective, the lattice Boltzmann (LB) method has recently been demonstrated to be a very effective numerical approach for the representation of many complex fluid systems.

 The LB method, which originated from the lattice gas automata, formally derives its basis from kinetic formulations that represent the streaming of the particle distribution functions and followed by their collisions. Here, the streaming is represented as a perfect-shift advection along lattice links, whereas the effects of collision are modeled as a relaxation process, while obeying appropriate conservation laws. The emergent continuum fluid behavior, then, arises as the averaged effect of such streaming and local collision steps. As such, the LB method, which may be characterized as a mesoscopic approach and derived as a minimal kinetic equation from the discretization of the Boltzmann equation~\cite{He1997}, has certain important physical and computational advantages. These include its natural ability to incorporate kinetic models for complex flows, ease of representation of boundary conditions, and inherent parallelization capabilities due to its localized computational steps facilitating efficient simulation of large problems. Naturally, the LB method has found a range of applications to a variety of complex flows, including multiphase flows, multicomponent systems, turbulence, particulate flows, thermal convective flows and microscale phenomena~\cite{Aidun2010,Succi2001,Guo2013,Kruger2016}. More recently, further improvements to the lattice Boltzmann method (LBM) has focused on enhancing its numerical stability, accuracy as well as computational efficiency.

 Based on how the collision step is modeled, both the numerical stability and accuracy are strongly influenced. A popular choice due to its simplicity is the single relaxation time (SRT) model, which represents the relaxation of the distribution functions to their equilibria at the same rate that represent the diffusive transport in fluids~\cite{Qian1992,Chen1992}. It has been shown to be prone to numerical instability for convection-dominated flows, have some limitations in the representation of boundary conditions, and are restricted in the simulation of thermal flows at a fixed Prandtl number. Multiple relaxation time (MRT) based LBM have been developed to address the above issues encountered with the SRT models~\cite{dHumieres2002}. In the MRT model, different raw moments, which are weighted summations of the product of the distribution function with the particle velocity components at different orders, are relaxed at different rates. Further improvements were more recently achieved by means of considering relaxation of central moments, which are obtained by shifting the particle velocity by the fluid velocity, to their local equilibria at different rates~\cite{Geier2006}. Such a central moment based MRT scheme is referred to as the cascaded MRT LBM. It representation in terms of a generalized local equilibrium was demonstrated by~\cite{Asinari2008}, construction of forcing terms and including them in 3D by~\cite{Premnath2009b} and a preconditioning formulation for convergence acceleration by~\cite{Hajabdollahi2017}. The significant advantages of the cascaded central moment LBM was recently demonstrated by~\cite{Geier2015}.

 During their early stages of development, the LB models focused on their applications to isothermal fluid flows. However, owing to numerous applications of thermal convection in fluids in natural settings and engineering, LB models for flows with heat transfer effects have also received considerable attention more recently. Generally, the following types of approaches have been considered in the LB framework for simulation of thermal convective flows: (a) Multispeed (MS) LB schemes~\cite{Alexander1993,Pavlo1998,McNamara1993,Chen1994}, (b) hybrid approach~(e.g.~\cite{Lallemand2003}), and (c) double distribution function (DDF) based LBM~\cite{He1998,Shi2004,Chai2014}. MS-Thermal LB models are obtained by including additional discrete velocities to the distribution function and using a higher order velocity expansion of the Maxwellian for modeling the equilibrium distribution; here, a single distribution function is used to represent the evolution of both velocity and temperature fields. Such approaches have severe restrictions in numerical stability and hence results in a narrow range of temperature variation. The hybrid approach considers using a LB model for the flow field and solves the thermal energy equation by means of another numerical scheme such as the finite difference method. The DDF-LB schemes considers the evolution of two different distribution functions, which have overcome many of the limitation of other formulations, and  are now  widely used.

 Most of the prior studies related to the development and applications of DDF-based LB models consider SRT models and generally limited to two-dimensions (2D)~\cite{Guo2002,Chopard2009,Chai2013}. The corresponding MRT based DDF-LB formulations were investigated by~\cite{Rasin2005,Yoshida2010,Wang2013,Chai2014}. For general practical applications, it is important to expand the capabilities of the LBM for thermal convective flows in 3D. However, only limited studies have so for been conducted in the literature in this regard. One of the earliest 3D LB models for heat transfer based on a passive scalar approach was presented by~\cite{Shan1997}, who performed simulations of Rayleigh-Benard convection using a SRT model. Subsequently, ~\cite{Peng2003} developed a 3D SRT LBM based on DDF approach and studied natural convection in a cubic cavity. More recently,~\cite{Yoshida2010} presented a MRT-LBE in 3D for CDE. Furthermore,~\cite{Du2011,Wu2011} and~\cite{Xia2014} employed the DDF-based LBM in 3D using the MRT formulation for certain heat transfer problems.

 In our present work, we present new 3D LB formulations based on the cascaded approach using central moments within a DDF approach to represent flows with thermal transport by convection and diffusion processes. Such a collision model is constructed using a moving frame of reference and involving central moment relaxation based on MRT. Due to the locality of the computational steps, these models maintain intrinsic parallelization properties enabling solution of large problems involving flows with heat transfer. Furthermore, the use of the cascaded central moment formulation would result in greater numerical stability to simulate 3D thermal convective flows. In this DDF approach, the cascaded LB scheme for the 3D fluid motion representing the solution of the Navier-Stokes equations (NSE) is based on a previous work. On the other hand, a new cascaded LB formulation for the solution of the 3D thermal transport equation represented by the CDE will be derived and investigated in this work. Here, it may be pointed out that the structure of the 3D cascaded collision operators to represent the CDE will be seen to be very different from that corresponding to the solution of the NSE using the same lattice. Such differences in the expressions for the collision kernels arise due to the number of collision invariants being different between solving the NSE (mass and momentum components, i.e. 1$+$3) and the CDE (scalar field, i.e. 1). In addition, in order to maintain generality of our 3D cascaded LB scheme, we consider representation of a local heat source in the CDE via a source term in the velocity space using a variable transformation. We will discuss derivations of the 3D cascaded LBE for the CDE representing the 3D thermal transport equations using both three-dimensional, fifteen velocity (D3Q15) lattice and its subset, viz., the three-dimensional, seven velocity (D3Q7) lattice. Finally, we present a quantitative validation of our 3D cascaded LB model for thermal convective flow by considering the simulation of 3D natural convection in a cubic cavity, which is a classical benchmark problem in this regard~\cite{Fusegi1991,Tric2000,Wakashima2004}. In particular, we will compare the structure of the velocity and temperature fields, as well as the heat transfer coefficient given in terms of the Nusselt number for different Rayleigh numbers, against the 3D benchmark solutions.

This paper is organized as follows. In the next section (Sec.2), we present the derivation of the 3D cascaded LBM for CDE representing the transport of the temperature field using the D3Q15 lattice following a brief exposition of the corresponding model for fluid flow. Section 3 presents the results and discussion of a numerical validation study involving the natural convection in a cubic enclosure containing air at different Rayleigh numbers. Finally, Sec.~4 provides a summary and conclusions arising from this work. In addition, the results of the derivation of the 3D cascaded LBM for CDE using a D3Q7 lattice are presented in Appendix C.

\section{\label{app:numerical}Three-dimensional Cascaded LBE for Thermal Convective Flows using D3Q15 Lattice}
A DDF-based cascaded LBM  for the computation of the coupled fluid motion with a scalar temperature field will now be constructed. Here, a distribution function $f_\alpha$, whose evolution is represented by a cascaded LB formulation for the solution of the Navier-Stokes equations (NSE), will be considered along with a separate distribution function $g_\alpha$, whose dynamics is represented by another cascaded LB scheme for the convection-diffusion equation (CDE) of the scalar field. To maintain generality, the fluid motion, i.e. velocity $\bm u$, is considered to be influenced by a spatially/temporally varying body force $\bm F$ and the scalar $\phi$ (such as the temperature $T$) by a local heat source $R$. We will derive the cascaded LB formulations for the typical lattice in 3D, i.e. the three-dimensional, fifteen velocity (D3Q15) lattice.

\subsection{\label{app:numerical}3D Cascaded LB Model for Fluid Flow}
Our goal is to first solve for the flow field represented by the 3D NSE given by
\begin{subequations}
 \begin{align}
&\partial_t \rho+\bm \nabla \cdot (\rho \bm u)=0, \label{eq:1a}\\
&\partial_t(\rho \bm u)+\bm \nabla\cdot(\rho \bm u \bm u)=-\bm \nabla P+\bm \nabla\cdot\mathbf \Pi_v +\bm F, \label{eq:1b}
\end{align}
\end{subequations}
where $\rho=\rho(\bm x,t)$ and $\bm u(\bm x,t)$ are the local fluid density and velocity, respectively, at a location $\bm x=(x,y,z)$ and time $t$. Here, $P, \mathbf \Pi_v$ and $\bm F$ represent the pressure, viscous stress tensor, and a local body force, respectively. It is assumed that $\bm F=\bm F(x,t)$. The 3D central moment LBM for the solution of Eqs.~(\ref{eq:1a}) and~(\ref{eq:1b}), including a local source term $S_\alpha$ in the velocity space for the D3Q15 lattice is presented in~\cite{Premnath2011three} as an extension of the cascaded LB model derived by~\cite{Geier2006}. A trapezoidal rule is considered in the characteristic integration of the source term to maintain second order accuracy, and then a variable transformation $\bar f_\alpha=f_\alpha-\frac{1}{2}S_\alpha$ is introduced to remove implicitness. Here, $\alpha = 0,1,\ldots, 14$. Briefly, the 3D cascaded LBM for fluid flow with a body force may then be written as~\cite{Premnath2011three}
\begin{subequations}
 \begin{align}
\tilde{\bar{f}}_{\alpha}(\bm{x},t)=\bar{f}_{\alpha}(\bm{x},t)+ (\tensor{K}\cdot\hat {\mathbf {g}})_{\alpha}+S_{\alpha}(\bm{x},t),\label{eq:2a}\\
\bar{f}_{\alpha}(\bm{x}+\bm{e}_{\alpha},t+1)=\tilde{\bar{f}}_{\alpha}(\bm{x},t). \label{eq:2b}
\end{align}
\end{subequations}

Here, Eqs.~(\ref{eq:2a}) and~(\ref{eq:2b}) represent the collision and streaming steps, respectively. $\tilde{\bar{f}}_{\alpha}$ represents the post-collision distribution function, $\tensor K$ is the orthogonal collision matrix, and $\hat {\mathbf {g}}$ is the collision kernel, which is obtained from the relaxation of the central moments at different orders to their corresponding local equilibria. While the focus here is on the derivation of a new cascaded LBE for the 3D CDE as discussed in what follows, for completeness, we present a summary of the expressions for $\tensor K$, $\hat {\mathbf {g}}$, $S_{\alpha}$, and $\tilde{\bar{f}}_{\alpha}$ for the solution of the 3D NSE in Appendix A. Once the distribution functions are updated, the hydrodynamic variables are obtained from the various kinetic moments as
\begin{equation}
\rho=\Sigma_{\alpha} {\bar f_{\alpha}}, \quad \rho \bm{ u}=\Sigma_{\alpha} \bar{f}_\alpha \bm {e}_\alpha+\frac{1}{2}\bm F. \label{eq:3}
\end{equation}

\subsection{\label{app:ChapmanEnskoganalysis1} 3D Cascaded LB Model for Transport of Temperature Field}
We now present a derivation of a 3D cascaded LBM on a D3Q15 lattice for the transport of any generic scalar field $\phi$ (such as temperature, i.e. $\phi=T$), which satisfies the following CDE:
\begin{equation}
 \partial_t\phi+\bm \nabla \cdot (\bm u \phi)=\bm \nabla \cdot(D_{\phi}\bm \nabla \phi)+R,
\label{eq:4}
\end{equation}
where $\phi=\phi(\bm x,t)$, ${\bm \nabla}=(\partial_x,\partial_y,\partial_z)$, $D_{\phi}$ is the coefficient of diffusivity, $R=R(\bm x,t)$ is the local source term, and the velocity field $\bm u$ can be obtained from the LB model discussed earlier.
The approach that is taken in this regard consists of the following overall steps:~(1) Construct an orthogonal moment basis starting from an initial set of linearly independent nonorthogonal basis vectors for the D3Q15 lattice.~(2) Prescribe expressions for the continuous central moments of equilibria and the source term at different orders and set them equal to their discrete central moments used in the cascaded LB formulation; obtain corresponding raw moments at different orders.~(3) Determine the structure of the cascaded collision kernel via considering a relaxation of central moments to their local equilibria at different orders, and obtain the source terms in the velocity space.\\

The components of the particle velocity for the D3Q15 lattice can be written as
\begin{eqnarray}
&\ket{e_{\alpha x}} =\left(     0,     1,    -1,     0,     0,  0,0,1,-1, 1, -1,1,-1,1,-1 \right)^\dag, \nonumber\\
&\ket{e_{\alpha y}} =\left(     0,     0,    0,     1,     -1,  0,0,1,1, -1, -1,1,1,-1,-1 \right)^\dag,
\nonumber\\
&\ket{e_{\alpha z}} =\left(     0,     0,    0,     0,     0,  1,-1,1,1, 1, 1,-1,-1,-1,-1 \right)^\dag,
\end{eqnarray}
and a corresponding unit vector may be represented by
\begin{eqnarray}
&\ket{\phi} =\left(     1,     1,    1,     1,     1,  1,1,1,1, 1, 1,1,1,1,1 \right)^\dag.
\label{eq:a1}
\end{eqnarray}
Here, we have used the notations $\bra{\cdot}$ and $\ket{\cdot}$ to represent the row and the column vectors respectively, $\dag$ is the transpose operator, and the operation $\braket{\bm a|\bm b}$ represents the dot product of any two vectors $\bm a$ and $\bm b$. Using successively higher order orders of the monomials $e^m_{\alpha x}e^n_{\alpha y} e^p_{\alpha z}$, we can write the following nonorthogonal basis vectors
\begin{eqnarray}
&&\ket{T_{0}}=\ket{\phi},\nonumber\\
&&\ket{T_{1}}=\ket{e_{\alpha x}},\quad
\ket{T_{2}}=\ket{e_{\alpha y}},\quad
\ket{T_{3}}=\ket{e_{\alpha z}},\nonumber\\
&&\ket{T_{4}}=\ket{e_{\alpha x}e_{\alpha y}},\quad
\ket{T_{5}}=\ket{e_{\alpha x}e_{\alpha z}},\quad
\ket{T_{6}}=\ket{e_{\alpha y}e_{\alpha z}},\nonumber\\
&&\ket{T_{7}}=\ket{e_{\alpha x}^2-e_{\alpha y}^2},\quad
\ket{T_{8}}=\ket{e_{\alpha x}^2-e_{\alpha z}^2},\quad
\ket{T_{9}}=\ket{e_{\alpha x}^2+e_{\alpha y}^2+e_{\alpha z}^2},\nonumber\\
&&\ket{T_{10}}=\ket{e_{\alpha x}(e_{\alpha x}^2+e_{\alpha y}^2+e_{\alpha z}^2)},\quad
\ket{T_{11}}=\ket{e_{\alpha y}(e_{\alpha x}^2+e_{\alpha y}^2+e_{\alpha z}^2)},\\
&&\ket{T_{12}}=\ket{e_{\alpha z}(e_{\alpha x}^2+e_{\alpha y}^2+e_{\alpha z}^2)},\quad
\ket{T_{13}}=\ket{e_{\alpha x}e_{\alpha y}e_{\alpha z}},\nonumber\\
&&\ket{T_{14}}=\ket{e_{\alpha x}^2e_{\alpha y}^2+e_{\alpha x}^2e_{\alpha z}^2+e_{\alpha y}^2e_{\alpha z}^2}.\nonumber
\end{eqnarray}

By applying the Gram-Schmidt orthogonalization method on the above set, we can obtain the corresponding set of orthogonal basis vectors, which are grouped together into the following collision matrix $\tensor{K}$ as
\begin{align}
&\tensor{K}=\left[{{\mathbf{K}}_{0}},{\mathbf{K}_{1}},{\mathbf{K}_{2}},{\mathbf{K}_{3}},{\mathbf{K}_{4}},{\mathbf{K}_{5}},{\mathbf{K}_{6}},{\mathbf{K}_{7}},{\mathbf{K}_{8}},{\mathbf{K}_{9}},{\mathbf{K}_{10}},{\mathbf{K}_{11}},{\mathbf{K}_{12}},{\mathbf{K}_{13}},{\mathbf{K}_{14}}\right]\nonumber\\
           \text{where}\nonumber\\
&\mathbf{K_{0}}=\ket{\phi},\nonumber\\
&\mathbf{K_{1}}=\ket{e_{\alpha x}},\quad
\mathbf{K_{2}}=\ket{e_{\alpha y}},\quad
\mathbf{K_{3}}=\ket{e_{\alpha z}},\nonumber\\
&\mathbf{K_{4}}=\ket{e_{\alpha x}e_{\alpha y}},\quad
\mathbf{K_{5}}=\ket{e_{\alpha x}e_{\alpha z}},\quad
\mathbf{K_{6}}=\ket{e_{\alpha y}e_{\alpha z}},\nonumber\\
&\mathbf{K_{7}}=\ket{e_{\alpha x}^2-e_{\alpha y}^2},\quad
\mathbf{K_{8}}=\ket{e_{\alpha x}^2+e_{\alpha y}^2+e_{\alpha z}^2}-3\ket{e_{\alpha z}^2},\quad
\mathbf{K_{9}}=\ket{e_{\alpha x}^2+e_{\alpha y}^2+e_{\alpha z}^2}-2\ket{\phi},\nonumber\\
&\mathbf{K_{10}}=5\ket{e_{\alpha x}(e_{\alpha x}^2+e_{\alpha y}^2+e_{\alpha z}^2)}-13\ket{e_{\alpha x}},\quad
\mathbf{K_{11}}=5\ket{e_{\alpha y}(e_{\alpha x}^2+e_{\alpha y}^2+e_{\alpha z}^2)}-13\ket{e_{\alpha y}},\nonumber\\
&\mathbf{K_{12}}=5\ket{e_{\alpha z}(e_{\alpha x}^2+e_{\alpha y}^2+e_{\alpha z}^2)}-13\ket{e_{\alpha z}},\nonumber\\
&\mathbf{K_{13}}=\ket{e_{\alpha x}e_{\alpha y}e_{\alpha z}},\nonumber\\
&\mathbf{K_{14}}=30\ket{e_{\alpha x}^2e_{\alpha y}^2+e_{\alpha x}^2e_{\alpha z}^2+e_{\alpha y}^2e_{\alpha z}^2}-40\ket{e_{\alpha x}^2+e_{\alpha y}^2+e_{\alpha z}^2}+32\ket{\phi}.
\end{align}
Then, we define the continuous central moments of equilibria needed in the construction of the cascaded collision kernel for the 3D CDE as follows:
\begin{equation}
\widehat{\Pi}^{{eq,\phi }}_{x^my^nz^p}=\int_{-\infty}^{\infty}\int_{-\infty}^{\infty}\int_{-\infty}^{\infty}g^{eq}(\xi_x-u_x)^m(\xi_y-u_y)^n(\xi_z-u_z)^p d\xi_xd\xi_yd\xi_z,\label{eq:6}
\end{equation}
where $g^{{eq}}$ is the equilibrium distribution function in the continuous velocity space ($\xi_x,\xi_y,\xi_z$) for the scalar field $\phi$, which is given by $g^{eq}\equiv
g^{eq}(\phi,\bm{u},\bm{\xi})=\frac{\phi}{2\pi c_{s\phi}^{3/2}}\exp\left[-\frac{\left(\bm{\xi}-\bm{u}\right)^2}{2c_{s\phi}^2}\right]$. Here $c_{s\phi}$ is a free parameter, which will be related to the desired coefficient of diffusivity $D_\phi$ later. Typically, we set $c^2_{s\phi}=\frac{1}{3}$, though it can be chosen to be at other values and different from that for the cascaded LB model for the flow field (see Appendix A). Moreover, $\bm u$ in  the above is the fluid velocity  which is obtained in the previous section. It may be noted that the above equilibrium distribution function is obtained from the local Maxwellian by replacing the density with the scalar field $\phi$ used in our DDF scheme. Then, rewriting the component of Eq.~(\ref{eq:6}) in the increasing order of moments as
\begin{eqnarray}
\widehat{\Pi}^{eq,\phi}_{0}&=&\phi, \nonumber\\
\widehat{\Pi}^{eq,\phi}_{x}&=&\widehat{\Pi}^{eq,\phi}_{y}=\widehat{\Pi}^{eq,\phi}_{z}=0, \nonumber\\
\widehat{\Pi}^{eq,\phi}_{xx}&=&\widehat{\Pi}^{eq,\phi}_{yy}=\widehat{\Pi}^{eq,\phi}_{zz}=c_{s\phi}^2\phi, \nonumber\\
\widehat{\Pi}^{eq,\phi}_{xy}&=&\widehat{\Pi}^{eq,\phi}_{xz}=\widehat{\Pi}^{eq,\phi}_{yz}=
\widehat{\Pi}^{eq,\phi}_{xyy}=\widehat{\Pi}^{eq,\phi}_{xxy}=\widehat{\Pi}^{eq,\phi}_{xxz}=
\widehat{\Pi}^{eq,\phi}_{xyz}=0, \nonumber\\
\widehat{\Pi}^{eq,\phi}_{xxyy}&=&c_{s\phi}^4\phi. \nonumber
\end{eqnarray}
Here, and henceforth, the use of hat over a symbol represents any quantity in the moment space. Similarly, the continuous central moments due to source term $R$ in Eq.~(\ref{eq:4}) may be defined as
\begin{equation}
\widehat{\Gamma}^{R}_{x^my^nz^p}=\int_{-\infty}^{\infty}\int_{-\infty}^{\infty}\int_{-\infty}^{\infty}\Delta g^R(\xi_x-u_x)^m(\xi_y-u_y)^n(\xi_z-u_z)^p d\xi_xd\xi_yd\xi_z.
\label{eq:8}
\end{equation}
where $\Delta g^R$ is the change in the distribution for the scalar field due to the source term. As the source term $R$ can only effect the lowest, i.e. zeroth central moment, the component of Eq.~(\ref{eq:8}) maybe written as
\begin{eqnarray}
&\widehat{\Gamma}^{R}_{0}=R, \nonumber\\
&\widehat{\Gamma}^{R}_{x}=\widehat{\Gamma}^{R}_{y}=\widehat{\Gamma}^{R}_{z}=
\widehat{\Gamma}^{R}_{xx}=\widehat{\Gamma}^{R}_{yy}=\widehat{\Gamma}^{R}_{zz}=
\widehat{\Gamma}^{R}_{xy}=\widehat{\Gamma}^{R}_{xz}=\widehat{\Gamma}^{R}_{yz}=0, \nonumber\\
&\widehat{\Gamma}^{R}_{xyy}=\widehat{\Gamma}^{R}_{xxy}=\widehat{\Gamma}^{R}_{xxz}=
\widehat{\Gamma}^{R}_{xyz}=
\widehat{\Gamma}^{R}_{xxyy}=0.
\end{eqnarray}

The cascaded LBE representing the transport of the 3D CDE can be obtained by applying a trapezoidal rule for the treatment of the source term in the characteristic integration to maintain a second order accuracy. Thus, we have
\begin{equation}
g_{\alpha}(\bm{x}+\bm{e}_{\alpha},t+1)=g_{\alpha}(\bm{x},t)+\Omega_{\alpha}^{g}(\bm{x},t)+
\frac{1}{2}\left[S_{\alpha}^\phi(\bm{x},t)+S_{\alpha}^\phi(\bm{x}+\bm{e}_{\alpha},t+1)\right],
\label{eq:10}
\end{equation}
where $S_{\alpha}^\phi$ is the source term in the velocity space that effectively accounts for the term $R(\bm x,t)$ in the macroscopic CDE. In the above equation, the collision term $\Omega_{\alpha}^{g}(\bm{x},t)$ can be modeled by
\begin{equation}
\Omega_{\alpha}^{g}\equiv \Omega_{\alpha}^{g}(\mathbf{g},\mathbf{\widehat{h}})=(\tensor{K}\cdot \mathbf{\widehat{h}})_{\alpha},
\label{eq:11}
\end{equation}
where $\mathbf{g}=(g_0,g_1,\cdots, g_{14})^\dag$ is the vector of the distribution function in Eq.~(\ref{eq:10}), and $\mathbf{\widehat{h}}=(\widehat h_0,\widehat h_1,\cdots, \widehat h_{14})^\dag$ is the vector of the unknown collision kernel which will be determined later. For removing the implicitness, while maintaining a second-order accuracy, by applying a variable transformation~\cite{He1999}, $\bar{g}_{\alpha}=g_{\alpha}-\frac{1}{2}S_{\alpha}^\phi$ in Eq.~(\ref{eq:10}), we obtain
\begin{equation}
\bar{g}_{\alpha}(\bm{x}+\bm{e}_{\alpha},t+1)=\bar{g}_{\alpha}(\bm{x},t)+\Omega_{\alpha}^{g}(\bm{x},t)+
S_{\alpha}^\phi(\bm{x},t).
\label{eq:12}
\end{equation}

This 3D central moment LBE may be rewritten in terms of the following collision and streaming steps for the purpose of implementation as
\begin{subequations}
\begin{align}
\tilde{\bar{g}}_{\alpha}(\bm{x},t)=\bar{g}_{\alpha}(\bm{x},t)+ (\tensor{K}\cdot\hat {\mathbf {h}})_{\alpha}+S_{\alpha}^\phi(\bm{x},t),\label{eq:13a}\\
\bar{g}_{\alpha}(\bm{x}+\bm{e}_{\alpha},t+1)=\tilde{\bar{g}}_{\alpha}(\bm{x},t) \label{eq:13b},
\end{align}
\end{subequations}
where the symbol $\sim$ in the above represents the post collision distribution function. In order to construct the structure of the cascaded collision and the source terms for representing the 3D CDE, we first define the following set of discrete central moments as
\begin{eqnarray}
\left( \begin{array}{l}
{{\hat {\kappa}^\phi }_{{x^m}{y^n}{z^p}}}\\
\hat \kappa _{{x^m}{y^n}{z^p}}^{eq ,\phi}\\
{{\hat {\sigma}^\phi }_{{x^m}{y^n}{z^p}}}\\
{\hat{\bar {\kappa}}^\phi_{{x^m}{y^n}{z^p}}}
\end{array} \right) = \sum\limits_\alpha  {\left( \begin{array}{l}
{g_\alpha }\\
g_\alpha ^{eq}\\
{\kern 1pt} {S_{\alpha}^\phi }\\
{{\bar g}_\alpha }
\end{array} \right)} {({e_{\alpha x}} - {u_x})^m}{({e_{\alpha y}} - {u_y})^n} {({e_{\alpha z}} - {u_z})^p},\label{eq:14}
\end{eqnarray}
where ${\hat{ \bar {\kappa}}^\phi_{{x^m}{y^n}{z^p}}}={{\hat {\kappa}^\phi }_{{x^m}{y^n}{z^p}}}-\frac{1}{2}{{\hat {\sigma}^\phi }_{{x^m}{y^n}{z^p}}}$. Then, by equating the discrete central moments of the equilibrium distribution function and source term with their corresponding continuous central moments at different orders, i.e. $\hat \kappa  _{{x^m}{y^n}{z^p}}^{eq ,\phi}=\widehat{\Pi}^{eq,\phi}_{x^my^nz^p}$ and ${{\hat {\sigma}^\phi }_{{x^m}{y^n}{z^p}}}=\widehat{\Gamma}^{R}_{x^my^nz^p}$, respectively, we get
 \begin{eqnarray}
\hat \kappa _{0}^{eq ,\phi}&=&\phi,\nonumber \\
\hat \kappa _{x}^{eq ,\phi}&=&\hat \kappa _{x}^{eq ,\phi}=\hat \kappa _{x}^{eq ,\phi}=0,\nonumber \\
\hat \kappa _{xx}^{eq ,\phi}&=&\hat \kappa _{yy}^{eq ,\phi}=\hat \kappa _{zz}^{eq ,\phi}=c_{s\phi}^2\phi,\nonumber \\
\hat \kappa _{xy}^{eq ,\phi}&=&\hat \kappa _{xz}^{eq ,\phi}=\hat \kappa _{yz}^{eq ,\phi}=
\hat \kappa _{xyy}^{eq ,\phi}=\hat \kappa _{xxy}^{eq ,\phi}=\hat \kappa _{xxz}^{eq ,\phi}=
\hat \kappa _{xyz}^{eq ,\phi}=0,\nonumber \\
\hat \kappa _{xxyy}^{eq ,\phi}&=&c_{s\phi}^4\phi, \label{eq:15}
\end{eqnarray}
and
\begin{align}
&\hat {\sigma} _{0}^{\phi}=R,\nonumber \\
&\hat {\sigma}_{x}^{\phi}=\hat {\sigma} _{y}^{\phi}=\hat {\sigma} _{z}^{\phi}=0,\nonumber \\
&\hat {\sigma} _{xx}^{\phi}=\hat {\sigma} _{yy}^{\phi}=\hat {\sigma} _{zz}^{\phi}=
\hat {\sigma} _{xy}^{\phi}=\hat {\sigma} _{xz}^{\phi}=\hat {\sigma} _{yz}^{\phi}=0,\nonumber \\
&\hat {\sigma} _{xyy}^{\phi}=\hat {\sigma}_{xxy}^{\phi}=\hat {\sigma} _{xxz}^{\phi}=
\hat {\sigma} _{xyz}^{\phi}=0,\nonumber \\
&\hat {\sigma} _{xxyy}^{\phi}=0 \label{eq:16}.
\end{align}
Since the calculations are effectively carried out in term of various raw moments, we define the following set of the raw moments at different orders  as
 \begin{eqnarray}
\left( \begin{array}{l}
{{\hat {\kappa}^{\phi'} }_{{x^m}{y^n}{z^p}}}\\
\hat \kappa _{{x^m}{y^n}{z^p}}^{eq ,{\phi'}}\\
{{\hat {\sigma}^{\phi'} }_{{x^m}{y^n}{z^p}}}\\
{\hat{\bar{\kappa}}^{\phi'}_{{x^m}{y^n}{z^p}}}
\end{array} \right) = \sum\limits_\alpha  {\left( \begin{array}{l}
{g_\alpha }\\
g_\alpha ^{eq}\\
{\kern 1pt} {S_{\alpha}^{\phi'} }\\
{{\bar g}_\alpha }
\end{array} \right)} e_{\alpha x}^m e_{\alpha y}^n e_{\alpha z}^p,\label{eq:17}
\end{eqnarray}
where $\hat{ \bar {\kappa}}^{\phi'}_{{x^m}{y^n}{z^p}}={{\hat {\kappa}^{\phi'} }_{{x^m}{y^n}{z^p}}}-\frac{1}{2}{{\hat {\sigma}^{\phi'} }_{{x^m}{y^n}{z^p}}}$, and the use of primes over any symbol here and henceforth refer to raw moments. From the above, we first determine the expressions for the source terms in the velocity space in Eq.~(\ref{eq:12}). In this regard, as an intermediate step, by applying the binomial theorem on Eq.~(\ref{eq:16}), we obtain the discrete raw moments of the source terms at different orders as
 \begin{eqnarray*}
&\hat {\sigma} _{0}^{\phi'}=R,\nonumber \\
&\hat {\sigma}_{x}^{\phi'}=u_xR,\quad \hat {\sigma} _{y}^{\phi'}=u_y R,\quad \hat {\sigma} _{z}^{\phi'}=u_z R,\nonumber \\
&\hat {\sigma}_{xx}^{\phi'}=u_x^2R,\quad \hat {\sigma} _{yy}^{\phi'}=u_y^2 R,\quad\hat {\sigma} _{zz}^{\phi'}=u_z^2 R,\nonumber \\
&\hat {\sigma}_{xy}^{\phi'}=u_xu_yR,\quad \hat {\sigma} _{xz}^{\phi'}=u_xu_z R,\quad\hat {\sigma} _{yz}^{\phi'}=u_yu_z R,\nonumber \\
&\hat {\sigma} _{xyy}^{\phi'}=u_xu_y^2R,\quad\hat {\sigma}_{xxy}^{\phi'}=u_x^2u_y R,\quad\hat {\sigma} _{xxz}^{\phi'}=u_x^2u_z R,\nonumber \\
&\hat {\sigma} _{xyz}^{\phi'}=u_xu_yu_zR,\nonumber \\
&\hat {\sigma} _{xxyy}^{\phi'}=u_x^2u_y^2R\nonumber .\label{eq:centralmomentforcing8}
\end{eqnarray*}

Next, from this, we obtain the source terms projected to the orthogonal basis vector $\tensor K$, i.e. $\widehat{\bm{m}}^{s,\phi}=(\tensor K_\alpha\cdot \tensor S^\phi)$, where  $\tensor S^\phi=(S^\phi_0,S^\phi_1,S^\phi_2,\dots,S^\phi_{14})$. That is,
\begin{align}
\widehat{m}^{s,\phi}_{0}&=\braket{K_0|S_{\alpha}^\phi}=R,\qquad
\widehat{m}^{s,\phi}_{1}=\braket{K_1|S_{\alpha}^\phi}=u_xR, \qquad
\widehat{m}^{s,\phi}_{2}=\braket{K_2|S_{\alpha}^\phi}=u_yR,\nonumber \\
\widehat{m}^{s,\phi}_{3}&=\braket{K_3|S_{\alpha}^\phi}=u_zR, \qquad
\widehat{m}^{s,\phi}_{4}=\braket{K_4|S_{\alpha}^\phi}=u_xu_yR,\qquad
\widehat{m}^{s,\phi}_{5}=\braket{K_5|S_{\alpha}^\phi}=u_xu_zR, \nonumber\\
\widehat{m}^{s,\phi}_{6}&=\braket{K_6|S_{\alpha}^\phi}=u_yu_zR,\qquad
\widehat{m}^{s,\phi}_{7}=\braket{K_7|S_{\alpha}^\phi}=(u_x^2-u_y^2)R,\nonumber\\
\widehat{m}^{s,\phi}_{8}&=\braket{K_8|S_{\alpha}^\phi}=(u_x^2+u_y^2-2u_z^2)R,\qquad
\widehat{m}^{s,\phi}_{9}=\braket{K_9|S_{\alpha}}=(u_x^2+u_y^2+u_z^2-2)R, \nonumber\\
\widehat{m}^{s,\phi}_{10}&=\braket{K_{10}|S_{\alpha}^\phi}=5\left[(u_x^3+u_xu_y^2+u_xu_z^2)R)\right]-13u_xR, \nonumber\\
\widehat{m}^{s,\phi}_{11}&=\braket{K_{11}|S_{\alpha}^\phi}=5\left[(u_x^2u_y+u_y^3+u_yu_z^2)R)\right]-13u_yR, \nonumber\\
\widehat{m}^{s,\phi}_{12}&=\braket{K_{12}|S_{\alpha}^\phi}=5\left[(u_x^2u_z+u_y^2u_z+u_z^3)R)\right]-13u_zR, \nonumber\\
\widehat{m}^{s,\phi}_{13}&=\braket{K_{13}|S_{\alpha}^\phi}=u_xu_yu_zR, \nonumber\\
\widehat{m}^{s,\phi}_{14}&=\braket{K_{14}|S_{\alpha}^\phi}=30\left[(u_x^2u_y^2+u_x^2u_z^2+u_y^2u_z^2)R)\right]-40\left[(u_x^2+u_y^2+u_z^2)R)\right]+32R. \label{eq:19}
\end{align}
Finally, by inverting the above, i.e. $S^\phi=\tensor K^{-1}\cdot \widehat{\bm{m}}^{s,\phi}$, and exploiting the orthogonality of $\tensor K$, we can determine the explicit expressions for the source terms in the velocity space $S^\phi_\alpha$, which are listed in Appendix B.

In order to construct the collision kernel $\widehat {\mathbf h}$ for the 3D cascaded collision operator for the scalar field $\phi$, we need the raw moments of the collision kernel of different orders, i.e. ${\sum\limits_\alpha}(\tensor K\cdot\mathbf {\widehat{h}})_\alpha {e_{\alpha x}^m e_{\alpha y}^n e_{\alpha z}^p}$. Using the orthogonality property of $\tensor K$, and considering that the only conserved invariant of this 3D cascaded LBE is the scalar field $\phi$ corresponding to the zeroth moment (i.e. $\widehat{h}_0=0$), we get
\begin{eqnarray*}
&\sum_{\alpha}(\tensor{K}\cdot \mathbf{\widehat{h}})_{\alpha}
=0,\quad
\sum_{\alpha}(\tensor{K}\cdot \mathbf{\widehat{h}})_{\alpha}e_{\alpha x}=
10\widehat{h}_{1},\quad
\sum_{\alpha}(\tensor{K}\cdot \mathbf{\widehat{h}})_{\alpha}e_{\alpha y}=
10\widehat{h}_{2},\nonumber\\
&\sum_{\alpha}(\tensor{K}\cdot \mathbf{\widehat{h}})_{\alpha}e_{\alpha z}=
10\widehat{h}_{3},\quad
\sum_{\alpha}(\tensor{K}\cdot \mathbf{\widehat{h}})_{\alpha}e_{\alpha x}e_{\alpha y}=
8\widehat{h}_{4},\quad
\sum_{\alpha}(\tensor{K}\cdot \mathbf{\widehat{h}})_{\alpha}e_{\alpha x}e_{\alpha z}=
8\widehat{h}_{5},\nonumber\\
&\sum_{\alpha}(\tensor{K}\cdot \mathbf{\widehat{h}})_{\alpha}e_{\alpha y}e_{\alpha z}=
8\widehat{h}_{6},\quad
\sum_{\alpha}(\tensor{K}\cdot \mathbf{\widehat{h}})_{\alpha}e_{\alpha x}^2=2\widehat{h}_{7}+2\widehat{h}_{8}+6\widehat{h}_{9},\nonumber\\
&\sum_{\alpha}(\tensor{K}\cdot \mathbf{\widehat{h}})_{\alpha}e_{\alpha y}^2=-2\widehat{h}_{7}+2\widehat{h}_{8}+6\widehat{h}_{9},\quad
\sum_{\alpha}(\tensor{K}\cdot \mathbf{\widehat{h}})_{\alpha}e_{\alpha z}^2=-4\widehat{h}_{8}+6\widehat{h}_{9},\nonumber\\
&\sum_{\alpha}(\tensor{K}\cdot \mathbf{\widehat{h}})_{\alpha}e_{\alpha x}e_{\alpha y}^2=16\widehat{h}_{10}+8\widehat{h}_{1},\quad
\sum_{\alpha}(\tensor{K}\cdot \mathbf{\widehat{h}})_{\alpha}e_{\alpha x}e_{\alpha z}^2=16\widehat{h}_{10}+8\widehat{h}_{1},\nonumber\\
&\sum_{\alpha}(\tensor{K}\cdot \mathbf{\widehat{h}})_{\alpha}e_{\alpha x}^2e_{\alpha y}=16\widehat{h}_{11}+8\widehat{h}_{2},\quad
\sum_{\alpha}(\tensor{K}\cdot \mathbf{\widehat{h}})_{\alpha}e_{\alpha y}e_{\alpha z}^2=16\widehat{h}_{11}+8\widehat{h}_{2},\nonumber\\
&\sum_{\alpha}(\tensor{K}\cdot \mathbf{\widehat{h}})_{\alpha}e_{\alpha x}^2e_{\alpha z}=16\widehat{h}_{12}+8\widehat{h}_{3},\quad
\sum_{\alpha}(\tensor{K}\cdot \mathbf{\widehat{h}})_{\alpha}e_{\alpha y}^2e_{\alpha z}=16\widehat{h}_{12}+8\widehat{h}_{3},\nonumber\\
&\sum_{\alpha}(\tensor{K}\cdot \mathbf{\widehat{h}})_{\alpha}e_{\alpha x}e_{\alpha y}e_{\alpha z}=8\widehat{h}_{13},\quad
\sum_{\alpha}(\tensor{K}\cdot \mathbf{\widehat{h}})_{\alpha}e_{\alpha x}^2e_{\alpha y}^2=8\widehat{h}_{9}+16\widehat{h}_{14},\nonumber\allowdisplaybreaks\\
&\sum_{\alpha}(\tensor{K}\cdot \mathbf{\widehat{h}})_{\alpha}e_{\alpha x}^2e_{\alpha z}^2=8\widehat{h}_{9}+16\widehat{h}_{14},\quad
\sum_{\alpha}(\tensor{K}\cdot \mathbf{\widehat{h}})_{\alpha}e_{\alpha y}^2e_{\alpha z}^2=8\widehat{h}_{9}+16\widehat{h}_{14}.
\end{eqnarray*}
At this point, it is important to highlight the significant difference in the derivation of the cascaded LBE for the fluid velocity $\bm u$ (given in Appendix A) and that for the scalar field $\phi$ considered here. In the case of the fluid flow, the mass and momentum components are the conserved variables for collision, and hence its corresponding collision kernel components will be zero, i.e. $\widehat{g}_0=\widehat{g}_1=\widehat{g}_2=\widehat{g}_3=0$. However, in the present case, only the zeroth moment, i.e. the passive scalar field is the conserved moment during collision. Hence, $\widehat h_0=0$, but $\widehat h_1\neq\widehat h_2\neq\widehat h_3\neq 0$. Due to these differences, it will be evident in the following that the expressions for the cascaded collision operator for the scalar field $\phi$ are quite different from those for the  fluid velocity $\bm u$ given in Appendix A.

Finally based on the above, we determine the structure of the 3D cascaded collision operator for the scalar field $\phi$ satisfying the CDE as follows: Beginning first at the lowest order non-conserved post-collision central moments, i.e. those for the first order moment components here, we set them equal to their corresponding equilibrium states as an intermediate step. When the tentative expression for a particular collision kernel component $\widehat h_\alpha$($\alpha	\geqslant 1$) is obtained in this manner, we discard the equilibrium assumption and multiply it by a corresponding relaxation parameter $\omega_\alpha^\phi$. This step allows for a relaxation process
in terms of the central moments to represent the effect of collision in the 3D cascaded LBM~\cite{Geier2006,Premnath2011three}. After considerable algebraic manipulations and simplifications, and using the notation
\begin{equation}
\widehat{\overline{\eta}}_{x^my^nz^p}^{\phi'}=\widehat{\overline{\kappa}}_{x^my^nz^p}^{\phi'}+\widehat{{\sigma}}_{x^my^nz^p}^{\phi'}
\label{eq:21}
\end{equation}
for brevity, we summarize the final expressions for the collision kernel components $\widehat h _\alpha$ as
\begin{eqnarray}
\widehat{h}_0&=&0,\nonumber\\
\widehat{h}_1&=&\frac{\omega_1^\phi}{10}\left[\phi u_x-\widehat{\overline{\kappa}}_{x}^{\phi'}-u_x (R/2) \right],\nonumber\\
\widehat{h}_2&=&\frac{\omega_2^\phi}{10}\left[\phi u_y-\widehat{\overline{\kappa}}_{y}^{\phi'}-u_y (R/2) \right],\nonumber\\
\widehat{h}_3&=&\frac{\omega_3^\phi}{10}\left[\phi u_z-\widehat{\overline{\kappa}}_{z}^{\phi'}-u_z (R/2) \right],\nonumber\\
\widehat{h}_4&=&\frac{\omega_4^\phi}{8}\left[-\widehat{\overline{\eta}}_{xy}^{\phi'}+u_y\widehat{\overline{\eta}}_{x}^{\phi'}+u_x\widehat{\overline{\eta}}_{y}^{\phi'}-\left(\phi+\frac{R}{2}\right)u_xu_y\right]+\frac{5}{4}\left(u_y\widehat{h}_1+u_x\widehat{h}_2\right),\nonumber\\
\widehat{h}_5&=&\frac{\omega_5^\phi}{8}\left[-\widehat{\overline{\eta}}_{xz}^{\phi'}+u_z\widehat{\overline{\eta}}_{x}^{\phi'}+u_x\widehat{\overline{\eta}}_{z}^{\phi'}-\left(\phi+\frac{R}{2}\right)u_xu_z\right]+\frac{5}{4}\left(u_z\widehat{h}_1+u_x\widehat{h}_3\right),\nonumber\\
\widehat{h}_6&=&\frac{\omega_6^\phi}{8}\left[-\widehat{\overline{\eta}}_{yz}^{\phi'}+u_z\widehat{\overline{\eta}}_{y}^{\phi'}+u_y\widehat{\overline{\eta}}_{z}^{\phi'}-\left(\phi+\frac{R}{2}\right)u_yu_z\right]+\frac{5}{4}\left(u_z\widehat{h}_2+u_y\widehat{h}_3\right), \nonumber\\
\widehat{h}_7&=&\frac{\omega_7^\phi}{4}\left[-(\widehat{\overline{\eta}}_{xx}^{\phi'}-\widehat{\overline{\eta}}_{yy}^{\phi'})+2(u_x\widehat{\overline{\eta}}_{x}^{\phi'}-u_y\widehat{\overline{\eta}}_{y}^{\phi'})-\left(\phi+\frac{R}{2}\right)(u_x^2-u_y^2)\right]+ \nonumber\\
                                      &&5\left(u_x\widehat{h}_1-u_y\widehat{h}_2\right),\nonumber \\
\widehat{h}_8&=&\frac{\omega_8^\phi}{12}\left[-(\widehat{\overline{\eta}}_{xx}^{\phi'}+\widehat{\overline{\eta}}_{yy}^{\phi'}-2\widehat{\overline{\eta}}_{zz}^{\phi'})+2(u_x\widehat{\overline{\eta}}_{x}^{\phi'}+u_y\widehat{\overline{\eta}}_{y}^{\phi'}-2u_z\widehat{\overline{\eta}}_{z}^{\phi'})-\left(\phi+\frac{R}{2}\right)(u_x^2+u_y^2-2u_z^2)\right]+ \nonumber\\
                                      &&\frac{5}{3}\left(u_x\widehat{h}_1+u_y\widehat{h}_2-2u_z\widehat{h}_3\right),\nonumber \\
\widehat{h}_9&=&\frac{\omega_9^\phi}{18}\left[3c_{s\phi}^2\phi-(\widehat{\overline{\eta}}_{xx}^{\phi'}+\widehat{\overline{\eta}}_{yy}^{\phi'}+\widehat{\overline{\eta}}_{zz}^{\phi'})+2(u_x\widehat{\overline{\eta}}_{x}^{\phi'}+u_y\widehat{\overline{\eta}}_{y}^{\phi'}+u_z\widehat{\overline{\eta}}_{z}^{\phi'})-\left(\phi+\frac{R}{2}\right)(u_x^2+u_y^2+u_z^2)\right]+ \nonumber\\
                                      &&\frac{10}{9}\left(u_x\widehat{h}_1+u_y\widehat{h}_2+u_z\widehat{h}_3\right),\allowdisplaybreaks \nonumber \\
 \widehat{h}_{10}&=&\frac{\omega_{10}^\phi}{16}\left[-\widehat{\overline{\eta}}_{xyy}^{\phi'}+2u_y\widehat{\overline{\eta}}_{xy}^{\phi'}-u_y^2\widehat{\overline{\eta}}_{x}^{\phi'}
 +u_x\widehat{\overline{\eta}}_{yy}^{\phi'}-2u_xu_y\widehat{\overline{\eta}}_{y}^{\phi'}+\left(\phi+\frac{R}{2}\right)u_xu_y^2 \right]\nonumber \\&&-\left(\frac{5}{8}u_y^2+\frac{1}{2}\right)\widehat{h}_1-\frac{5}{4}
 u_xu_y\widehat{h}_2+u_y\widehat{h}_4+\frac{1}{8}u_x\left(-\widehat{h}_7+\widehat{h}_8+3\widehat{h}_9\right),\nonumber\\
 \widehat{h}_{11}&=&\frac{\omega_{11}^\phi}{16}\left[-\widehat{\overline{\eta}}_{xxy}^{\phi'}+2u_x\widehat{\overline{\eta}}_{xy}^{\phi'}+u_y\widehat{\overline{\eta}}_{xx}^{\phi'}-u_x^2\widehat{\overline{\eta}}_{y}^{\phi'}-2u_xu_y\widehat{\overline{\eta}}_{x}^{\phi'}
 +\left(\phi+\frac{R}{2}\right)u_x^2u_y\right]\nonumber\\&&-\frac{5}{4}u_xu_y \widehat{h}_{1}-\left(\frac{5}{8}u_x^2+\frac{1}{2}\right)\widehat{h}_2+u_x \widehat{h}_{4}+
 \frac{1}{8}u_y \left(\widehat{h}_{7}+\widehat{h}_{8}+3\widehat{h}_{9}\right),\nonumber\\
 \widehat{h}_{12}&=&\frac{\omega_{12}^\phi}{16}\left[-\widehat{\overline{\eta}}_{xxz}^{\phi'}+2u_x
\widehat{\overline{\eta}}_{xz}^{\phi'}+u_z
\widehat{\overline{\eta}}_{xx}^{\phi'}-2u_xu_z
\widehat{\overline{\eta}}_{x}^{\phi'}-u_x^2
\widehat{\overline{\eta}}_{z}^{\phi'}+\left(\phi+\frac{R}{2}\right)u_x^2u_z
 \right]\nonumber\\ &&-\frac{5}{4}u_xu_z\widehat{h}_{1}-\left(\frac{5}{8}u_x^2+\frac{1}{2}\right)\widehat{h}_{3}+u_x\widehat{h}_{5}+\frac{1}{8}u_z\left(\widehat{h}_{7}+\widehat{h}_{8}+3\widehat{h}_{9}\right),\nonumber\\
 \widehat{h}_{13}&=&\frac{\omega_{13}^\phi}{8}\left[-\widehat{\overline{\eta}}_{xyz}^{\phi'}+u_x\widehat{\overline{\eta}}_{yz}^{\phi'}+u_y\widehat{\overline{\eta}}_{xz}^{\phi'}
 +u_z\widehat{\overline{\eta}}_{xy}^{\phi'}-u_yu_z\widehat{\overline{\eta}}_{x}^{\phi'}-u_xu_z\widehat{\overline{\eta}}_{y}^{\phi'}
 -u_xu_y\widehat{\overline{\eta}}_{z}^{\phi'}+\left(\phi+\frac{R}{2}\right)u_xu_yu_z
 \right]\nonumber\\ &&-\frac{5}{4}\left(u_yu_z\widehat{h}_{1}+u_xu_z\widehat{h}_{2}+u_xu_y\widehat{h}_{3}\right)+u_z\widehat{h}_{4}+u_y\widehat{h}_{5}+u_x\widehat{h}_{6},\nonumber\\
 \widehat{h}_{14}&=&\frac{\omega_{14}^\phi}{16}\left[-\widehat{\overline{\eta}}_{xxyy}^{\phi'}+2u_y\widehat{\overline{\eta}}_{xxy}^{\phi'}+2u_x\widehat{\overline{\eta}}_{xyy}^{\phi'}
 -u_y^2\widehat{\overline{\eta}}_{xx}^{\phi'}-u_x^2\widehat{\overline{\eta}}_{yy}^{\phi'}-4u_xu_y\widehat{\overline{\eta}}_{xy}^{\phi'}+2u_xu_y^2\widehat{\overline{\eta}}_{x}^{\phi'}+2u_x^2
 u_y\widehat{\overline{\eta}}_{y}^{\phi'}+c_{s\phi}^4\phi\right.\nonumber\\ &&\left.-\left(\phi+\frac{R}{2}\right)u_x^2u_y^2\right]+
 \frac{5}{4}\left(u_xu_y^2\widehat{h}_{1}+u_x^2u_y\widehat{h}_{2}\right)
 -2u_xu_y\widehat{h}_{4}+\frac{1}{8}\left(u_x^2-u_y^2\right)\widehat{h}_{7}-\frac{1}{8}\left(u_x^2+u_y^2\right)\widehat{h}_{8}\nonumber \\&&
 -\left(\frac{3}{8}(u_x^2+u_y^2)+\frac{1}{2}\right)\widehat{h}_{9}+2\left(u_x\widehat{h}_{10}+u_y\widehat{h}_{11}\right)+u_y\widehat{h}_{2}+u_x\widehat{h}_{1}.
\end{eqnarray}
Here, the relaxation parameters $\omega_\alpha^\phi$, where $\alpha=1,2,\cdots 14$, satisfy the usual bounds $0< \omega_\alpha^\phi<2$. The above cascaded collision kernel represents the 3D convection-diffusion equation for any scalar field $\phi$ (such as temperature) with a source term, where the coefficient of diffusivity $D_\phi$ is related to the relaxation times of the first order moments by
\begin{equation}
D_\phi=c_{s\phi}^2\left(\frac{1}{\omega_j^\phi}-\frac{1}{2}\right),\quad j=1,2,3.
\label{eq:23}
\end{equation}
The remaining relaxation parameters for the higher order moments influence numerical stability and can be tuned independently. In this work, we set them to unity. Notice that the structure of the collision kernel of the 3D cascaded LBE for the scalar field $\phi$ is markedly different from that for the fluid flow (see Appendix A). In particular, the "cascaded" structure for the scalar field starts from the  second order moment components onward, while that for the fluid flow begins from the third order moments owing to the differences in the number of collision invariants as mentioned earlier.

Finally, by expanding the elements of the product $(\tensor K \cdot\widehat {\mathbf h})_\alpha$ in Eq.~({\ref{eq:11}}) and using it in Eq.~({\ref{eq:13a}}), the post-collision distribution functions in the velocity space, i.e. $\widetilde{\overline{g}}_{\alpha}$ are given by
\begin{eqnarray}
\widetilde{\overline{g}}_{0}&=&\overline{g}_{0}+\left[\widehat{h}_0-2\widehat{h}_9+32\widehat{h}_{14}\right]+S^\phi_0, \nonumber\\
\widetilde{\overline{g}}_{1}&=&\overline{g}_{1}+\left[\widehat{h}_0+\widehat{h}_1+\widehat{h}_{7}+\widehat{h}_{8}-\widehat{h}_{9}-8\widehat{h}_{10}-8\widehat{h}_{14}\right]+S^\phi_1,\nonumber\\
\widetilde{\overline{g}}_{2}&=&\overline{g}_{2}+\left[\widehat{h}_0-\widehat{h}_1+\widehat{h}_{7}+\widehat{h}_{8}-\widehat{h}_{9}+8\widehat{h}_{10}-8\widehat{h}_{14}\right]+S^\phi_2,\nonumber\\
\widetilde{\overline{g}}_{3}&=&\overline{g}_{3}+\left[\widehat{h}_0+\widehat{h}_2-\widehat{h}_{7}+\widehat{h}_{8}-\widehat{h}_{9}-8\widehat{h}_{11}-8\widehat{h}_{14}\right]+S^\phi_3,\nonumber\\
\widetilde{\overline{g}}_{4}&=&\overline{g}_{4}+\left[\widehat{h}_0-\widehat{h}_2-\widehat{h}_{7}+\widehat{h}_{8}-\widehat{h}_{9}+8\widehat{h}_{11}-8\widehat{h}_{14}\right]+S^\phi_4,\nonumber\\
\widetilde{\overline{g}}_{5}&=&\overline{g}_{5}+\left[\widehat{h}_0+\widehat{h}_3-2\widehat{h}_{8}-\widehat{h}_{9}-8\widehat{h}_{12}-8\widehat{h}_{14}\right]+S^\phi_5,\nonumber\\
\widetilde{\overline{g}}_{6}&=&\overline{g}_{6}+\left[\widehat{h}_0-\widehat{h}_3-2\widehat{h}_{8}-\widehat{h}_{9}+8\widehat{h}_{12}-8\widehat{h}_{14}\right]+S^\phi_6,\nonumber\\
\widetilde{\overline{g}}_{7}&=&\overline{g}_{7}+\left[\widehat{h}_0+\widehat{h}_1+\widehat{h}_2+\widehat{h}_3+\widehat{h}_4+\widehat{h}_5+\widehat{h}_6+\widehat{h}_9+2\widehat{h}_{10}+2\widehat{h}_{11}+2\widehat{h}_{12}\right.\nonumber\\
                                &&\left.+\widehat{h}_{13}+2\widehat{h}_{14}\right]+S^\phi_7,\nonumber\\
\widetilde{\overline{g}}_{8}&=&\overline{g}_{8}+\left[\widehat{h}_0-\widehat{h}_1+\widehat{h}_2+\widehat{h}_3-\widehat{h}_4-\widehat{h}_5+\widehat{h}_6+\widehat{h}_9-2\widehat{h}_{10}+2\widehat{h}_{11}+2\widehat{h}_{12}\right.\nonumber\\
                                &&\left.-\widehat{h}_{13}+2\widehat{h}_{14}\right]+S^\phi_8,\nonumber\\
\widetilde{\overline{g}}_{9}&=&\overline{g}_{9}+\left[\widehat{h}_0+\widehat{h}_1-\widehat{h}_2+\widehat{h}_3-\widehat{h}_4+\widehat{h}_5-\widehat{h}_6+\widehat{h}_9+2\widehat{h}_{10}-2\widehat{h}_{11}+2\widehat{h}_{12}\right.\nonumber\\
                                &&\left.-\widehat{h}_{13}+2\widehat{h}_{14}\right]+S^\phi_9,\nonumber\allowdisplaybreaks \\
\widetilde{\overline{g}}_{10}&=&\overline{g}_{10}+\left[\widehat{h}_0-\widehat{h}_1-\widehat{h}_2+\widehat{h}_3+\widehat{h}_4-\widehat{h}_5-\widehat{h}_6+\widehat{h}_9-2\widehat{h}_{10}-2\widehat{h}_{11}+2\widehat{h}_{12}\right.\nonumber\\
                                &&\left.+\widehat{h}_{13}+2\widehat{h}_{14}\right]+S^\phi_{10},\nonumber\\
\widetilde{\overline{g}}_{11}&=&\overline{g}_{11}+\left[\widehat{h}_0+\widehat{h}_1+\widehat{h}_2-\widehat{h}_3+\widehat{h}_4-\widehat{h}_5-\widehat{h}_6+\widehat{h}_9+2\widehat{h}_{10}+2\widehat{h}_{11}-2\widehat{h}_{12}\right.\nonumber\\
                                &&\left.-\widehat{h}_{13}+2\widehat{h}_{14}\right]+S^\phi_{11},\nonumber\\
\widetilde{\overline{g}}_{12}&=&\overline{g}_{12}+\left[\widehat{h}_0-\widehat{h}_1+\widehat{h}_2-\widehat{h}_3-\widehat{h}_4+\widehat{h}_5-\widehat{h}_6+\widehat{h}_9-2\widehat{h}_{10}+2\widehat{h}_{11}-2\widehat{h}_{12}\right.\nonumber\\
                                &&\left.+\widehat{h}_{13}+2\widehat{h}_{14}\right]+S^\phi_{12},\nonumber \\
\widetilde{\overline{g}}_{13}&=&\overline{g}_{13}+\left[\widehat{h}_0+\widehat{h}_1-\widehat{h}_2-\widehat{h}_3-\widehat{h}_4-\widehat{h}_5+\widehat{h}_6+\widehat{h}_9+2\widehat{h}_{10}-2\widehat{h}_{11}-2\widehat{h}_{12}\right.\nonumber\\
                                &&\left.+\widehat{h}_{13}+2\widehat{h}_{14}\right]+S^\phi_{13},\nonumber \allowdisplaybreaks \\
\widetilde{\overline{g}}_{14}&=&\overline{g}_{14}+\left[\widehat{h}_0-\widehat{h}_1-\widehat{h}_2-\widehat{h}_3+\widehat{h}_4+\widehat{h}_5+\widehat{h}_6+\widehat{h}_9-2\widehat{h}_{10}-2\widehat{h}_{11}-2\widehat{h}_{12}\right.\nonumber\\
                                &&\left.-\widehat{h}_{13}+2\widehat{h}_{14}\right]+S^\phi_{14}.
\end{eqnarray}
Upon performing the streaming step as given in Eq.~({\ref{eq:13b}}), using the above updated distribution function $\bar{g}_\alpha$, the scalar field $\phi$ can be finally computed as
\begin{equation}
\phi=\sum_{\alpha=0}^{14}\bar{g_\alpha}+\frac{1}{2}R.
\label{eq:25}
\end{equation}
For completeness, a simplified 3D cascaded LB formulation for the D3Q7 lattice is also presented in Appendix C.

\section{Results and Discussion}
A main objective of this section is to validate the new 3D thermal cascaded LB method discussed earlier for simulation of thermal convective flows. In this regard, we perform simulations of natural convection in a cubic cavity, based on which a comparison of the computed flow and thermal characteristic against 3D benchmark numerical solutions will be made. Natural convection of fluids in differentially heated enclosures has numerous engineering applications and arise in various natural settings. These include solar energy collectors, thermal energy storage systems, cooling of electronic devices, ventilation of buildings and crystal growth processes. It is chiefly characterized by the Rayleigh number representing the strength of the buoyancy effects relative to the counteracting  thermal and momentum diffusion effects, and the Prandtl number. Some of the classic 2D benchmark solutions for this problem include the results reported by~\cite{deVahl1983}. Given that the natural convective fluid motion in various cases of practical interest are three dimensional in nature, there have been considerable progress in obtaining benchmark numerical results for the 3D natural convection in a cubic enclosure~(e.g. \cite{Fusegi1991}) and our present study uses such data as part of the validation in the following.

A schematic of the geometric configuration for the physical model of the 3D cubic cavity considered and the coordinate system is shown in Fig.~1. It consists of a cubic enclosure of side length $L$ and the left wall and the right wall surfaces are maintained at temperatures of $T_L$ and $T_H$, respectively, where $T_H>T_L$; all the other four wall surfaces are maintained to be adiabatic. The convective fluid motion then arises naturally from the buoyancy force due to a local temperature difference with respect to a reference temperature in the presence of a gravity field. This thermally driven flow may be represented by means of the following body force $\bm F$ in the NSE in Eq.~({\ref{eq:1b}}) under the Boussinesq approximation as
\begin{equation}
\bm F=g\beta (T-T_0)\widehat{k},
\label{eq:23s}
\end{equation}
\begin{figure}[h!]
%\centering
\advance\leftskip-.1cm
\advance\rightskip 8cm
    \includegraphics[clip, trim=2cm 11cm 0.5cm 4cm, width=1.00\textwidth] {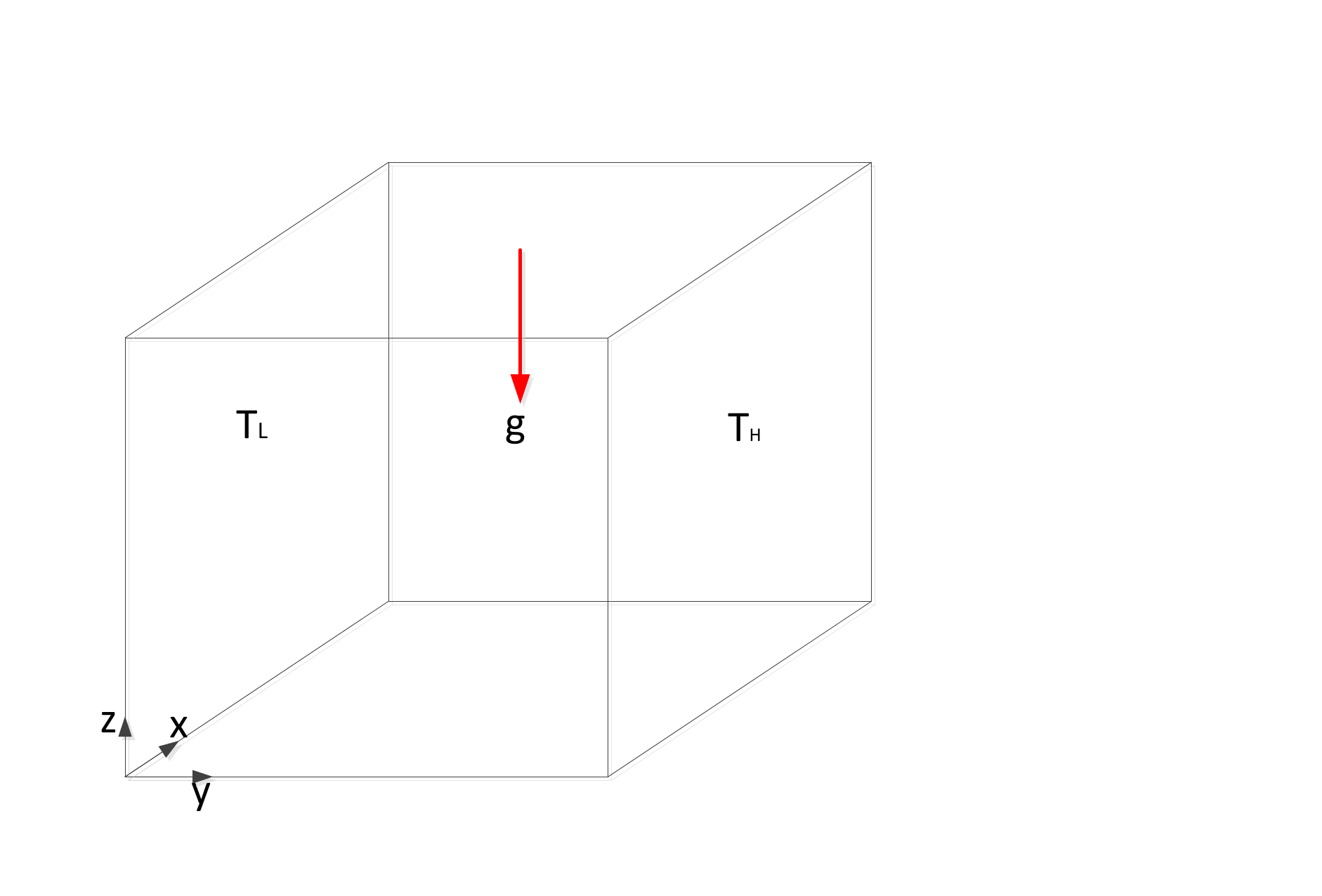}
\caption{ Geometric configuration for the physical model of the 3D cubic cavity and the coordinate system.}
\label{fig:schematic}
\end{figure}
where $\beta$ is the coefficient of thermal expansion, $T=T(x,y,z,t)$ is the local temperature field, $T_0=(T_L+T_H)/2$ is the referencee temperature, $g$  is the acceleration due to gravity, and $\widehat k$ is the unit vector in the positive $z$-direction in Fig.~1. This body force is used in the 3D cascaded LBM for fluid flow discussed in Sec.~2.1, while the local temperature field needed in Eq.~(\ref{eq:23s}) is obtained from the other 3D cascaded LBM for the thermal energy equation presented in Sec.~2.2. The velocity and the temperature boundary conditions may be summarized as
\begin{eqnarray}
&u_x=u_y=u_z=0\quad \text{for all walls}\\
&T(x,y=0,z)=T_L,\quad T(x,y=L,z)=T_H,\\
&\frac{\partial T}{\partial {\widehat {\bm n}}}=0 \quad\text{for all other walls}
\label{eq:23}
\end{eqnarray}
 where $\widehat {\bm n}$ is the wall normal direction. The standard half-way bounce back scheme is employed to implement the velocity boundary condition, and an anti-bounce back scheme is used to represent the Dirichlet boundary conditions for the scalar temperature field~\cite{Yoshida2010} and the Neumann boundary conditions are implemented using the scheme given in~\cite{Zhang2012}. The characteristic dimensionless Rayleigh number $\mbox{Ra}$ and the Prandtl number $\mbox{Pr}$ for this problem are given by
\begin{equation}
\mbox{Ra}=g\beta \Delta T L^3/(\alpha \nu),\quad \mbox{Pr}=\nu/\alpha,
\label{eq:23}
\end{equation}
where $\Delta T=T_H-T_L$ is the temperature difference between the hot and cold surface, $\alpha$ and $\nu$ are the thermal diffusivity and kinematic viscosity of the fluid, respectively. In the following, we will non-dimensionalize the coordinate lengths by the scale $L$, components of the velocity by $[g\beta L(T_H-T_L)]^{1/2}$ and the temperature by $T_0$. The corresponding dimensionless coordinates are then denoted by ($x,y,z$), the velocity field by $(u_x,u_y,u_z)$ and the temperature field by $T$. A key parameter characterizing the thermal transport during natural convection is the Nusselt number. The mean Nusselt number at either the hot or cold wall maybe represented as
 \begin{equation}
\mbox{Nu}_{mean}(z)=\int_{x=0}^1\frac{\partial{T{(x,y)}}}{\partial y}\Big|_{y=0 \text\ {or}\ y=1} dx.
\label{eq:23}
\end{equation}
In the following, we will consider simulations of natural convection of air ($\mbox{Pr}=0.71$) at different values of the Rayleigh number $\mbox{Ra}$. We will use the 3D  cascaded LBM based on the D3Q15 lattice in this regard, and using a grid resolution of $91\times 91\times 91$.

Figure~2 presents the temperature and velocity profiles between the adiabatic bottom and top walls in the $z$-direction ((a) and (c), respectively) and between the cold and hot surfaces in the $y-$direction ((b) and (d), respectively), along the symmetry plane ($x$=0.5) at $\mbox{Ra}=10^5$ computed using our 3D thermal cascaded LBM. Also, plotted in these subfigures in symbols are the prior reference benchmark solution based on directly solving the NSE~\cite{Fusegi1991}. It can be seen that the computed structure of both the temperature and velocity fields along different directions are in very good agrement with the benchmark numerical results. The slopes of the temperature fields near both the adiabatic and isothermal walls are found to be well captured by our 3D cascaded LBM based on the DDF formulation. Furthermore, from the velocity profiles shown in Figs.~2(c) and 2(d), it is evident, in particular, that both the peak magnitudes and their locations of the fluid convection velocity are well reproduced by our 3D cascaded LB model.
\begin{figure}[h!]
%\centering
\advance\leftskip-.9cm
\advance\rightskip 1.8cm
    \subfloat[]{
    \includegraphics[scale=0.5] {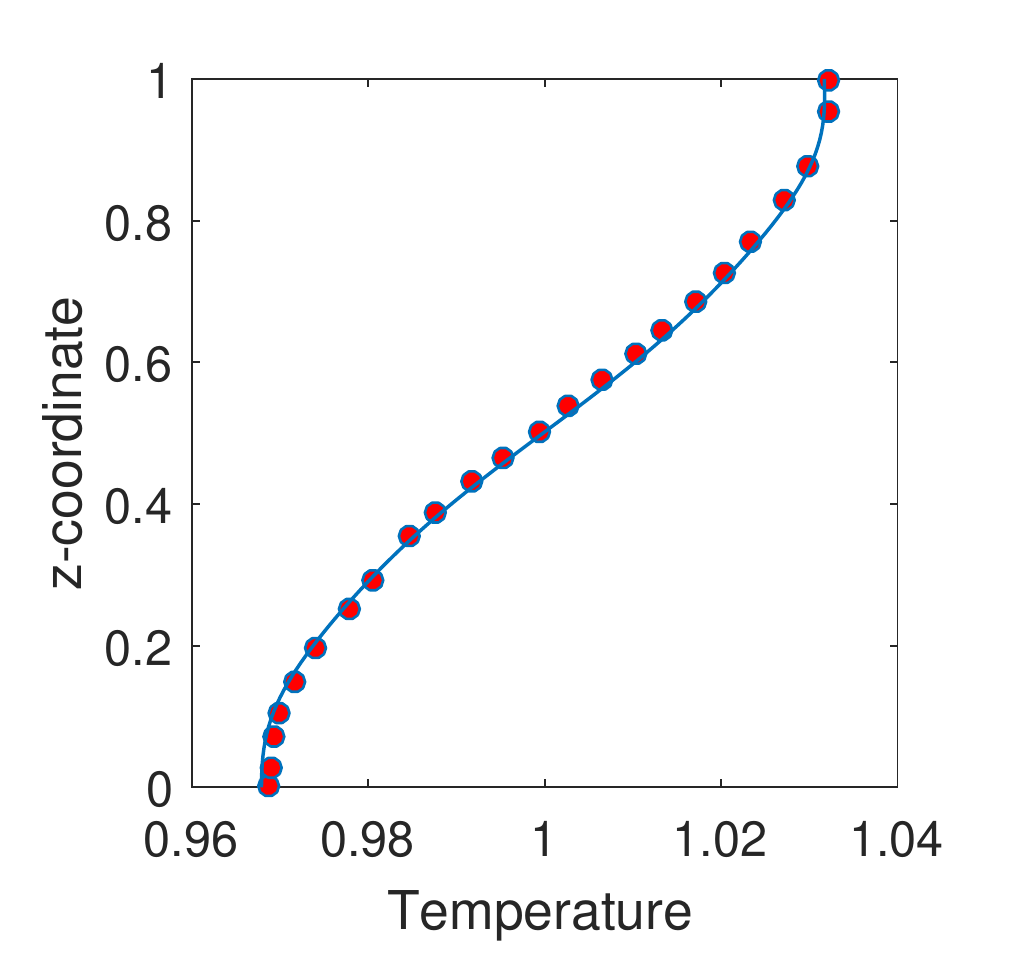}
        \label{fig:img1} } \hspace*{-1em}
    \hfill
    \subfloat[]{
      \includegraphics[scale=0.5] {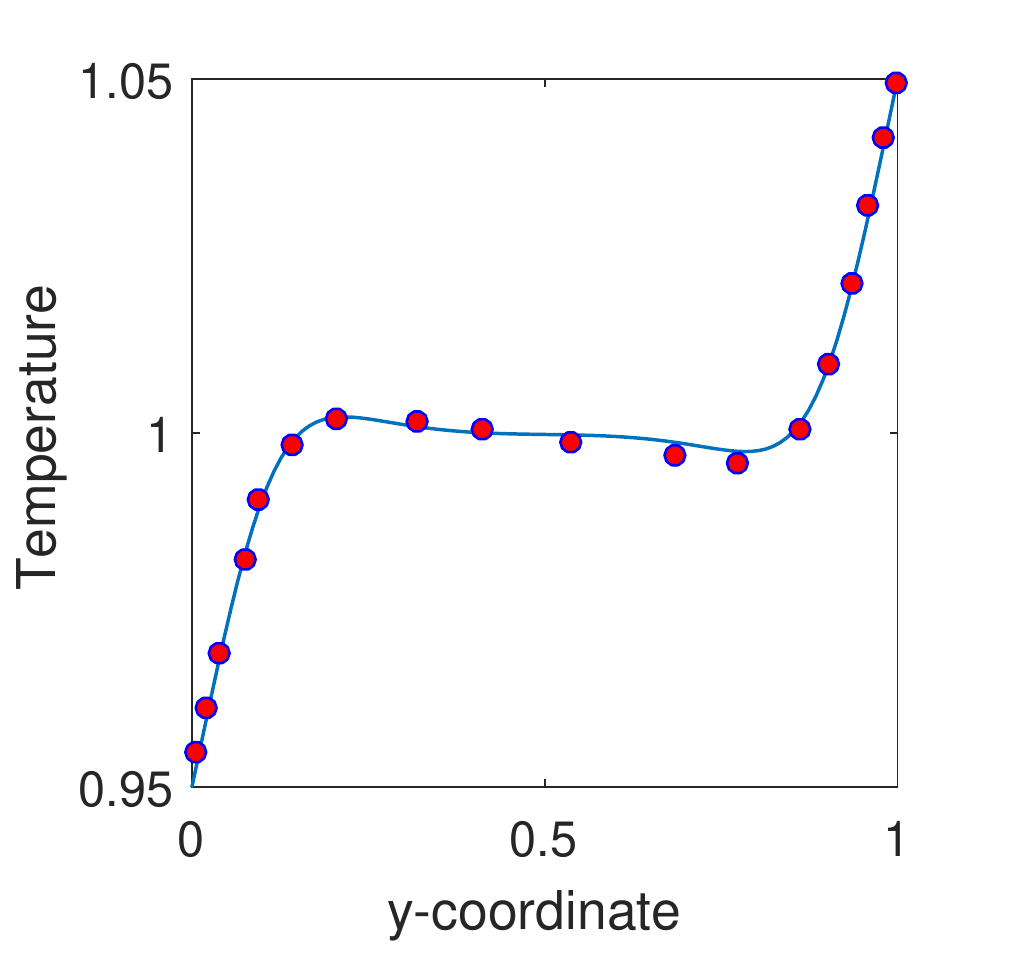}
        \label{fig:img2} } \\
%\advance\leftskip0cm
%\advance\rightskip -.62cm
             \subfloat[]{
        \includegraphics[scale=0.5] {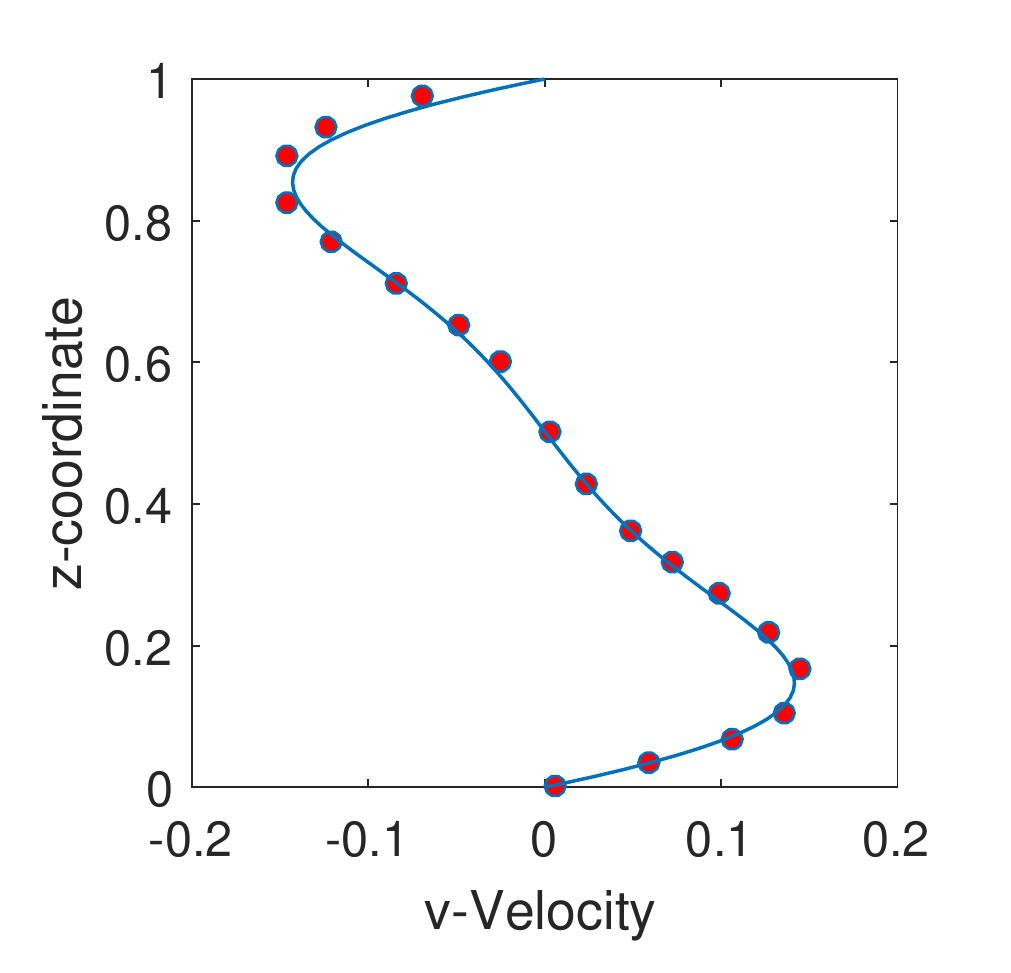}
        \label{fig:img3} } \hspace*{-1em}
     \hfill
        \subfloat[]{
          \includegraphics[scale=0.5] {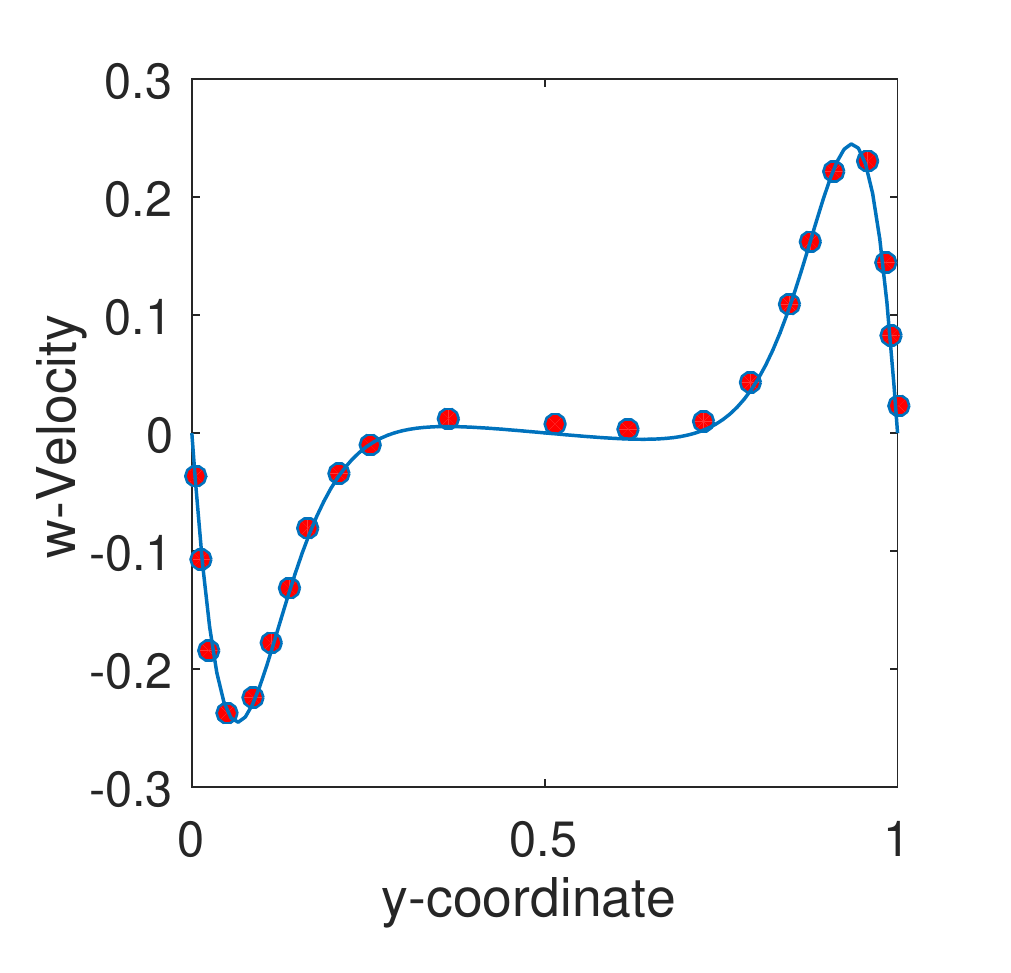}
          }
          \caption{Comparison of the temperature (top) and velocity profiles (bottom) for Rayleigh number $\mbox{Ra}=10^5$ on the symmetry ($x=0.5$) center plane $y-z$; symbols $``\circ"$ denote the reference benchmark solutions~\cite{Fusegi1991}, and lines $``-"$ represent results obtained using 3D thermal cascaded LBM.}
          \label{fig:vel}
\end{figure}

Figure 3 presents the distribution of streamlines arising due to natural convective fluid currents from differentially heated enclosures at two different Rayleigh numbers of $\mbox{Ra}=10^4$ and $Ra=10^5$ along center planes in different coordinate directions. In the vertical $y-z$ midplane ($x=0.5$), it can be seen that at lower $\mbox{Ra}$ of $10^4$, a central vortex appears as a dominant characteristic of the fluid motion. However, with increasing the Rayleigh number to $10^5$, when the natural convection effects becomes more pronounced, the central vortex breaks up into a set of two vortices. In addition, it is evident that there is a clustering of streamlines near the wall surfaces. Scale analysis predicts the boundary layer thickness $\delta$  near an isothermal wall set up by natural convection scales as $\mbox{Ra}^{-1/4}$. Hence, there is a thinner layer of fluid near walls that undergoes a more vigorous natural convection at higher $\mbox{Ra}$. On the $x-z$ midplane ($y=0.5$), in which side walls are adiabatic, it is seen that the heated fluid rises up, with the colder fluid moving down and being replaced by the heated one. The flow pattern is found to be three-dimensional in nature. On the $x-y$ midplane ($z=0.5$), near the hot wall, the fluid, which rises from the bottom of the cavity, moves towards the cold wall and after some distance changes the direction. For both the midplanes, as we increase the Rayleigh number, the thermal convective effects are found to be more dominant. These flow patterns are consistent with prior numerical solutions~\cite{Fusegi1991,Wakashima2004}.

 \begin{figure}[htbp]
\centering
%\advance\leftskip-1cm
%\advance\rightskip 1cm
    \subfloat{
    \includegraphics[width=5.5cm,scale=0.2] {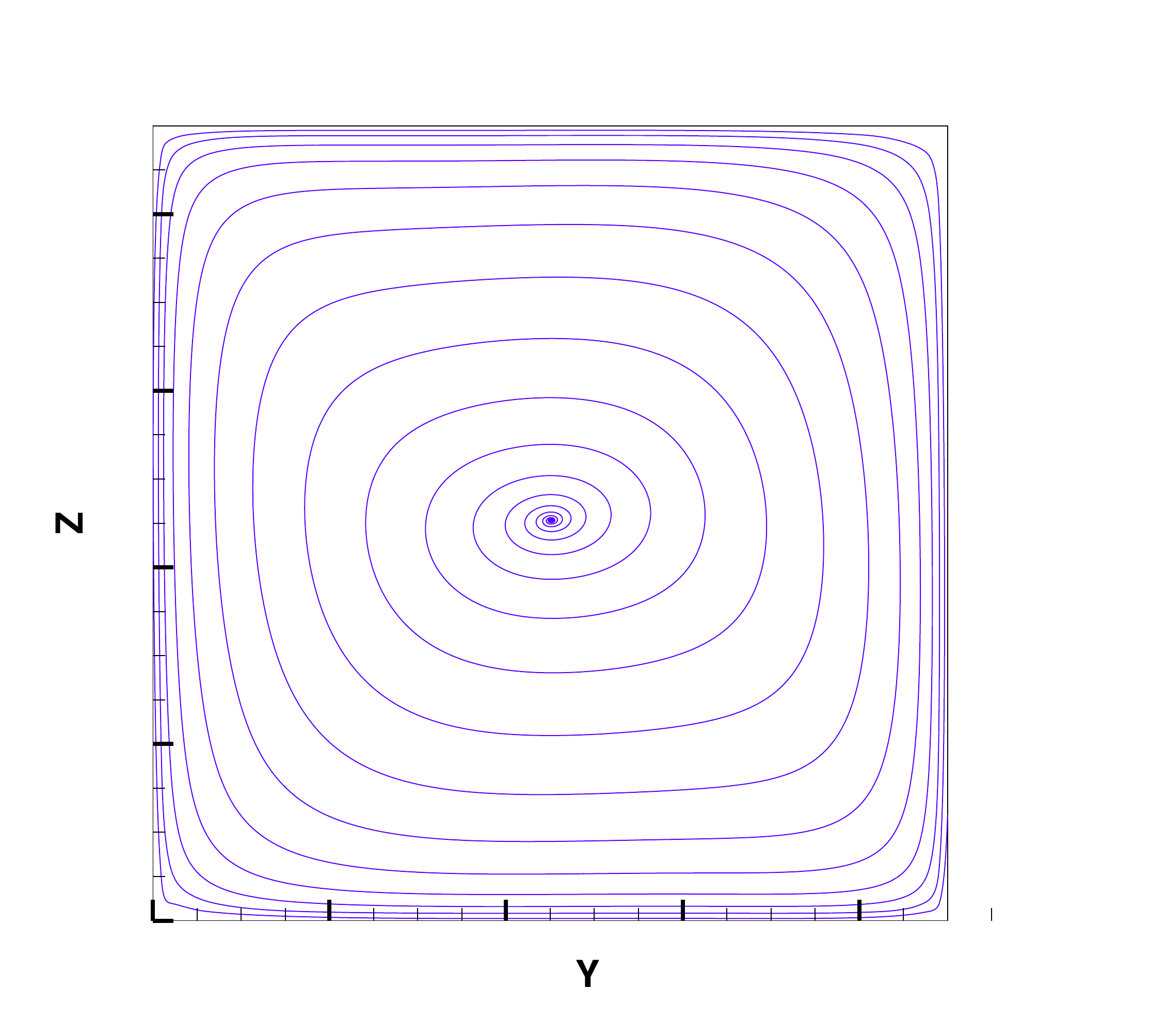}
        \label{fig:img1} } \hspace*{-35em}
    \hfill
    \subfloat{
      \includegraphics[width=5.5cm,scale=0.7] {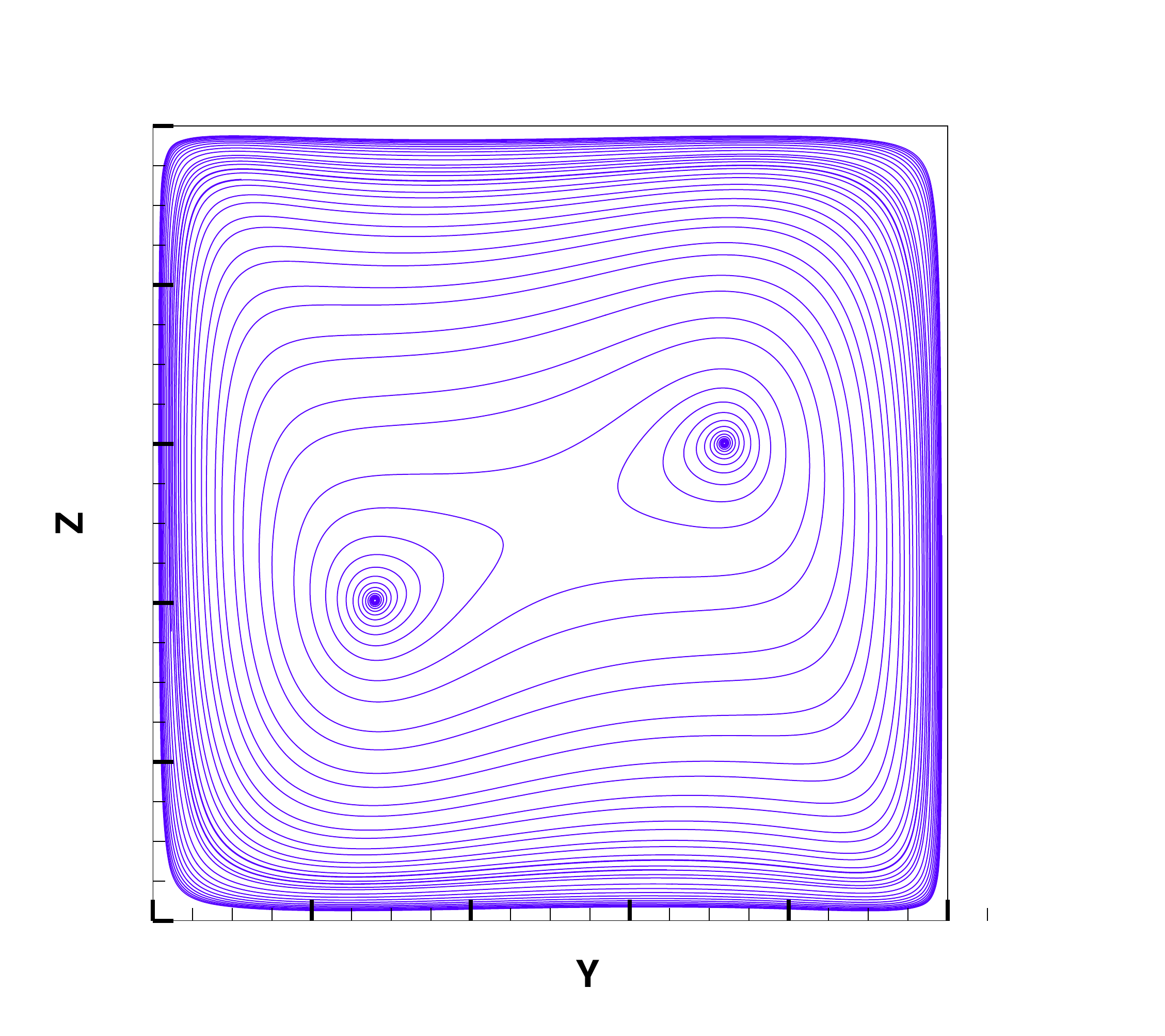}
        \label{fig:img2} } \\
%\advance\leftskip0cm
%\advance\rightskip -.62cm
             \subfloat{
        \includegraphics[width=5.5cm,scale=0.7] {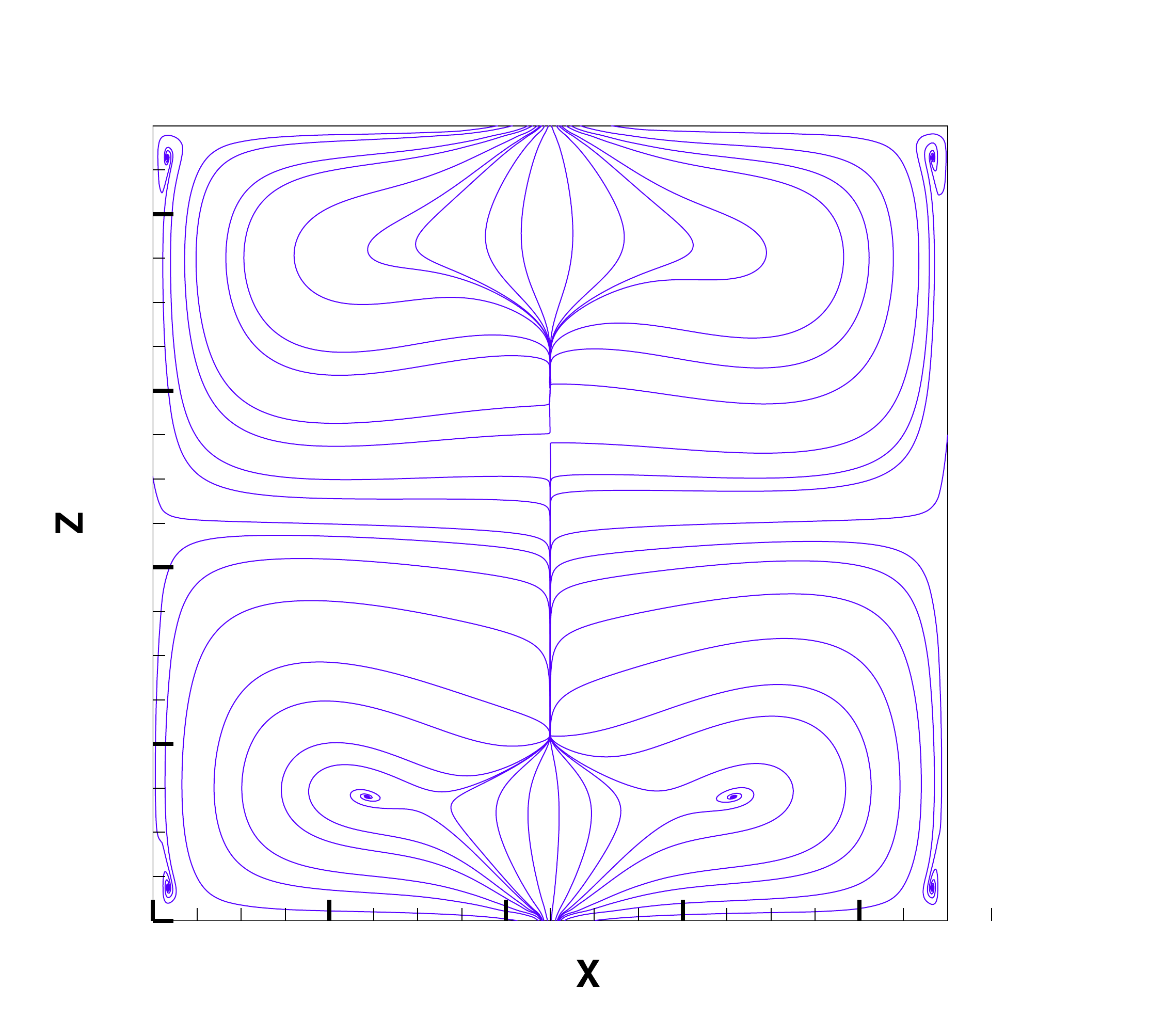}
        \label{fig:img3} } \hspace*{-35em}
     \hfill
        \subfloat{
          \includegraphics[width=5.5cm,scale=0.7]{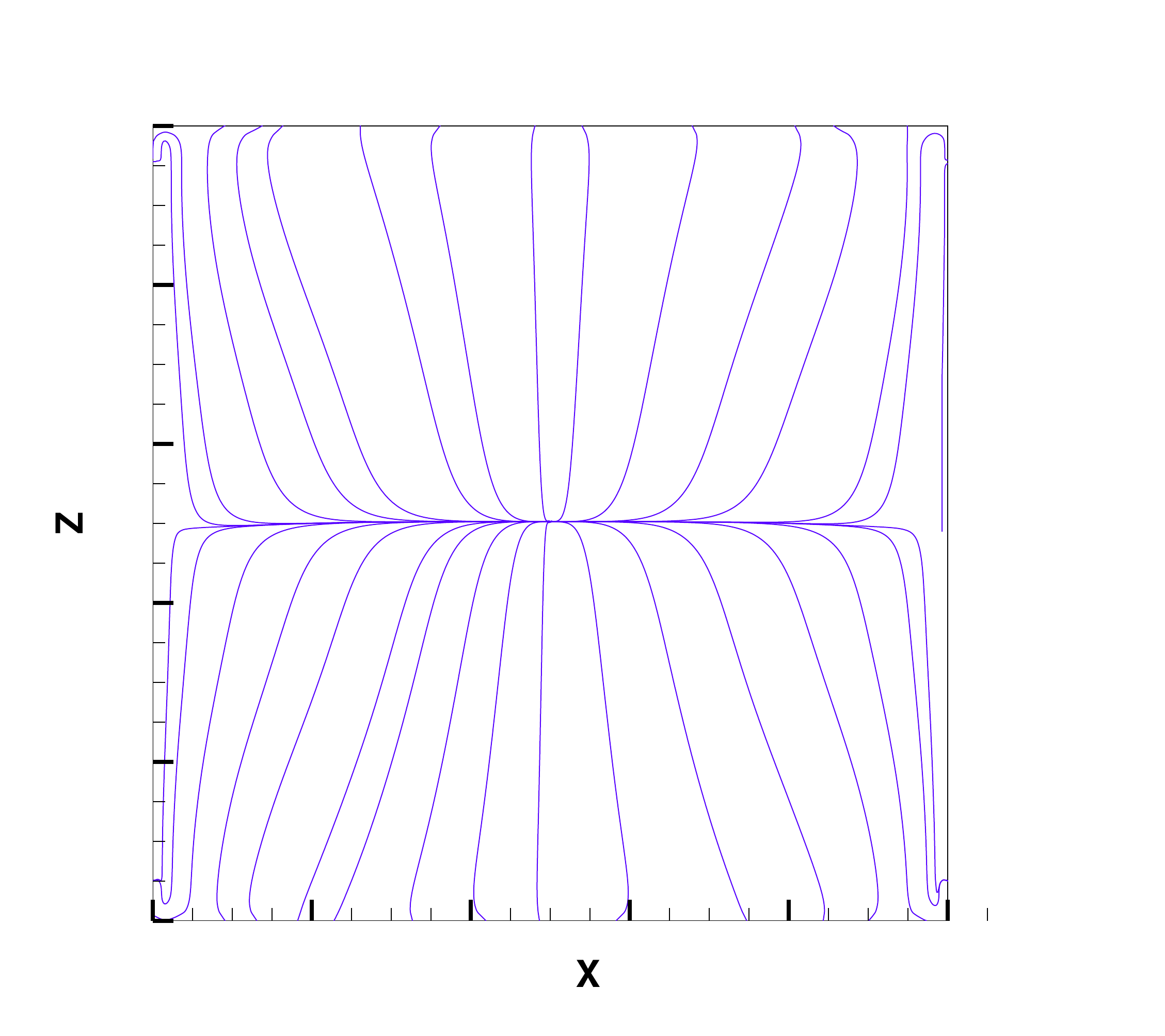}
        \label{fig:img4} }
              \\
              %\advance\leftskip-1.3cm
%\advance\rightskip -.63cm
       \subfloat{
       \includegraphics[width=5.5cm,scale=0.7]{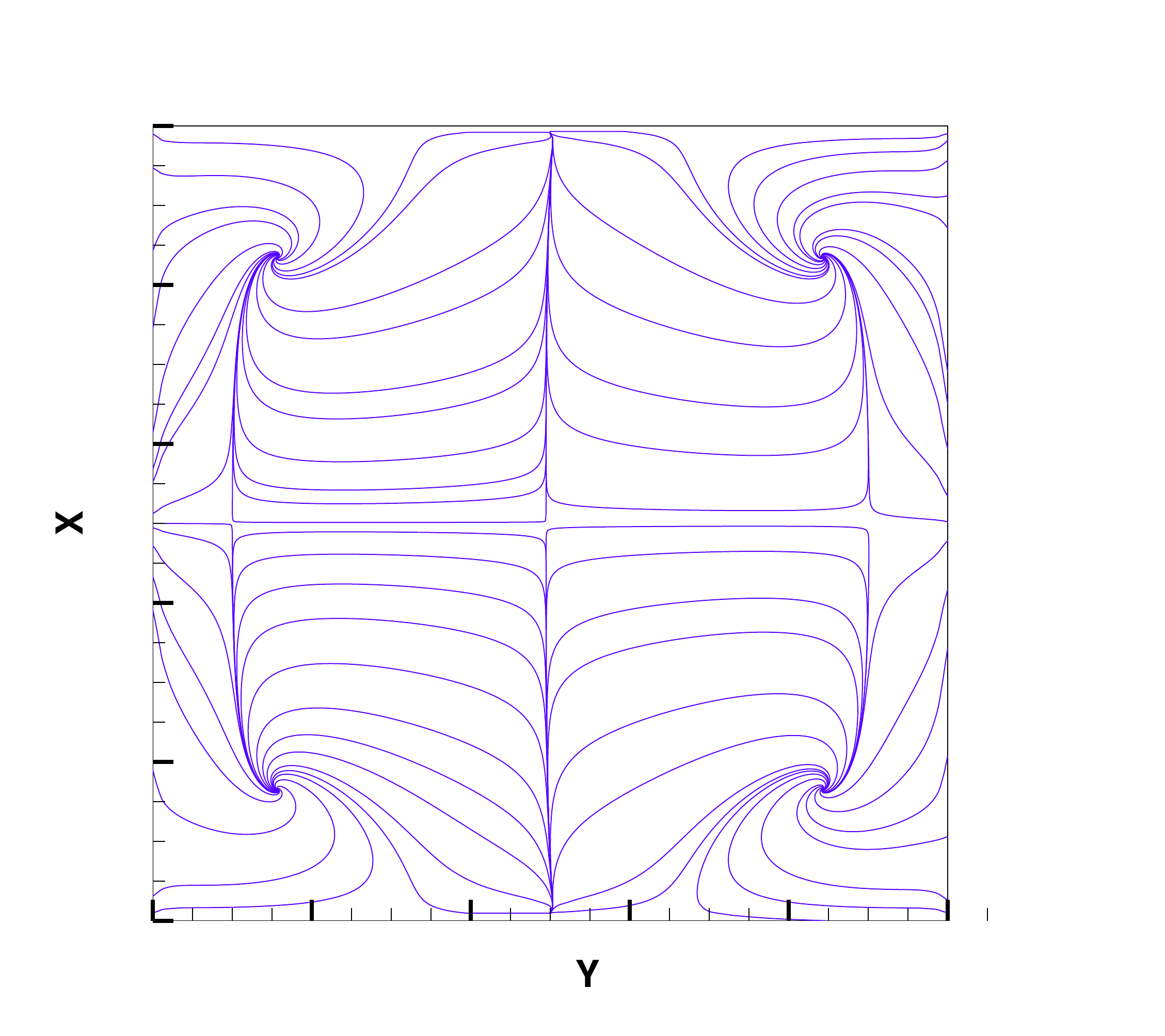}
        \label{fig:img5} }\hspace*{-35em}
             \subfloat{
         \includegraphics[width=5.5cm,scale=0.7]{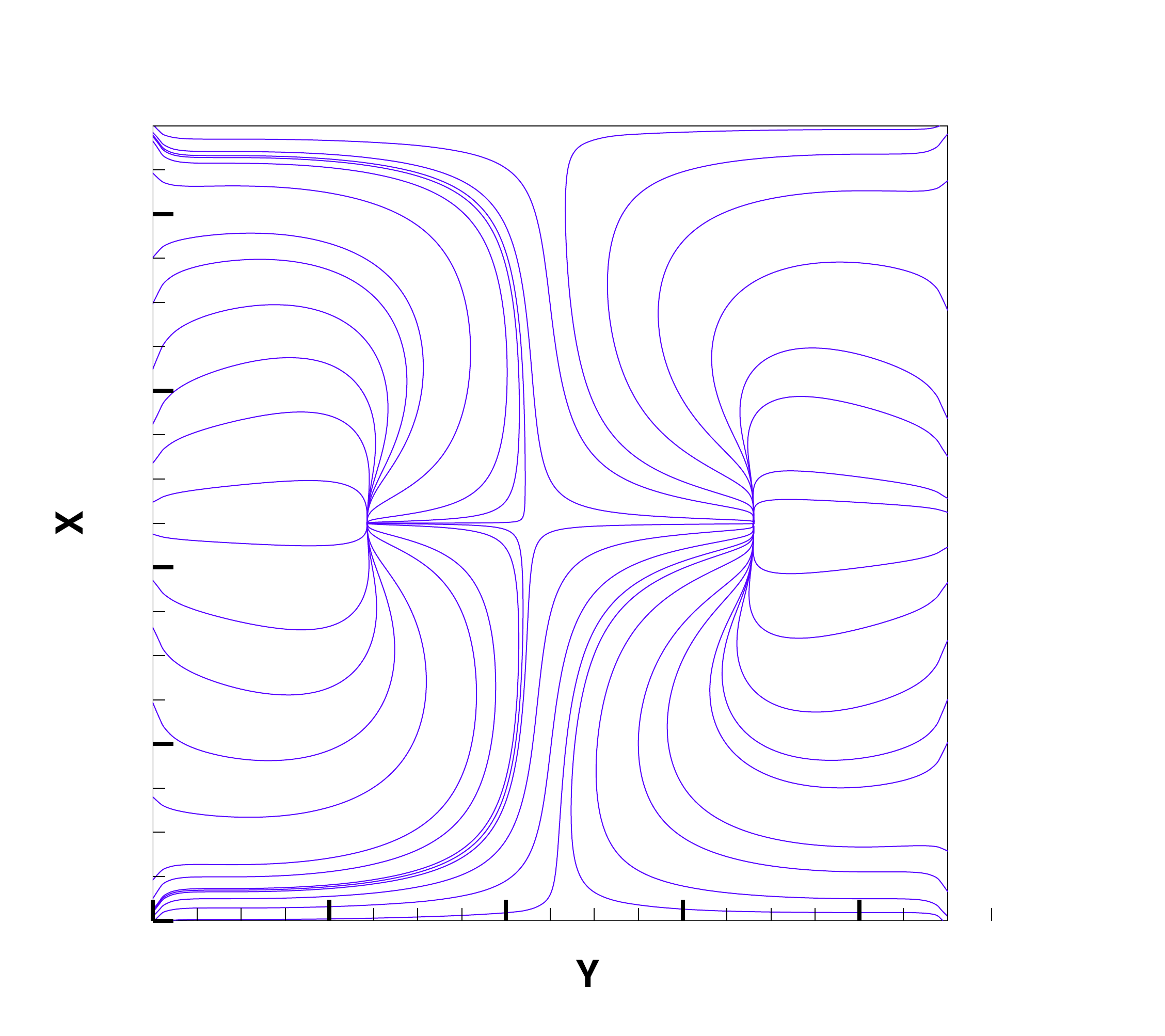}
        \label{fig:img6} }
        \caption{Projections of streamlines in natural convection in a 3D cavity computed using 3D thermal cascaded LBM along different center planes at Rayleigh numbers $\mbox{Ra}=10^4$ (left) and $\mbox{Ra}=10^5$ (right). Top row: $y-z$ plane, Middle row $x-z$ plane, Bottom row: $y-x$ plane.
    }
    \label{fig:str}
\end{figure}

\begin{figure}[htbp]
\centering
%\advance\leftskip-1cm
%\advance\rightskip 1cm
    \subfloat{
    \includegraphics[width=5.5cm,scale=0.2] {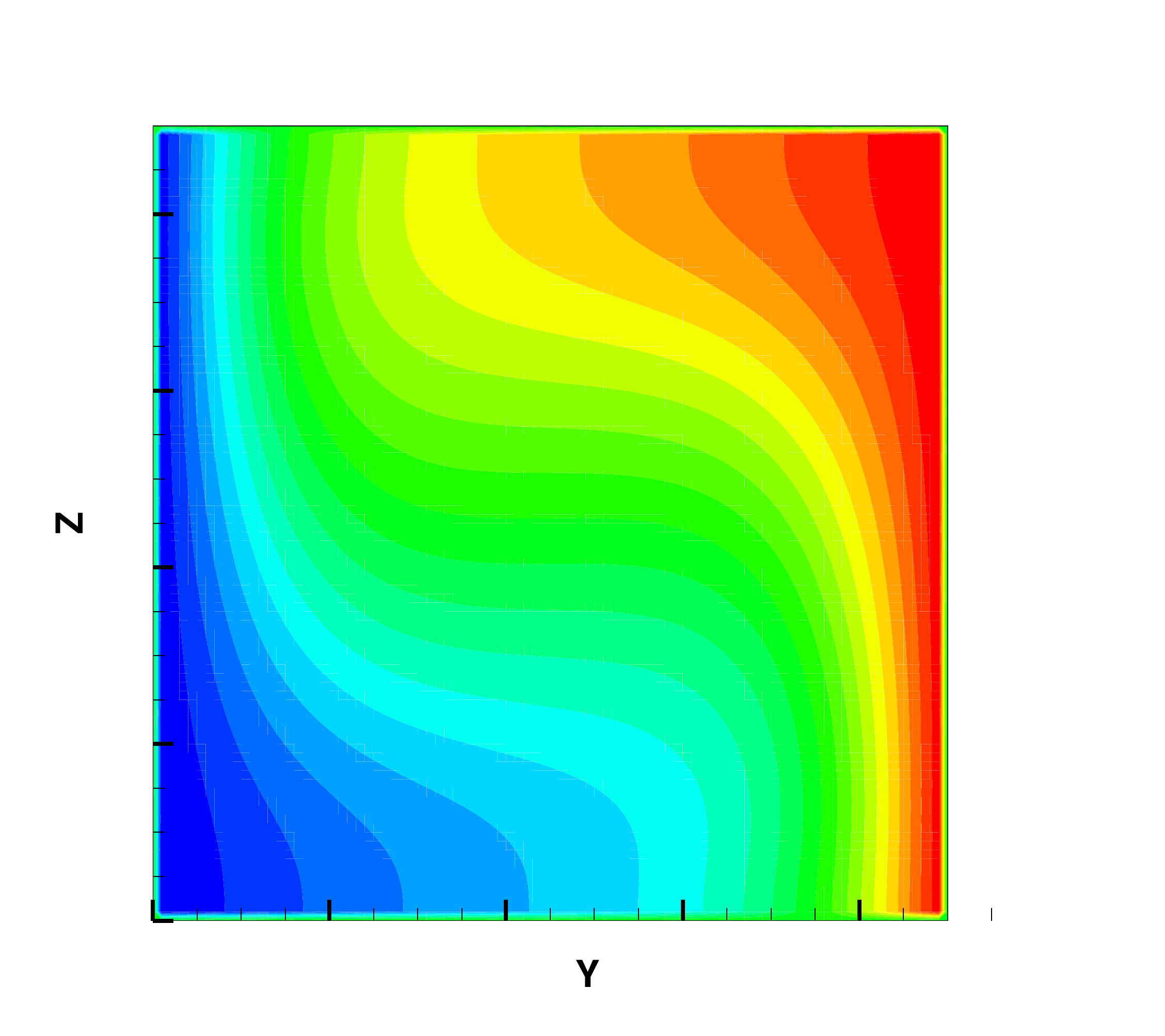}
        \label{fig:img1} } \hspace*{-35em}
    \hfill
    \subfloat{
      \includegraphics[width=5.5cm,scale=0.7] {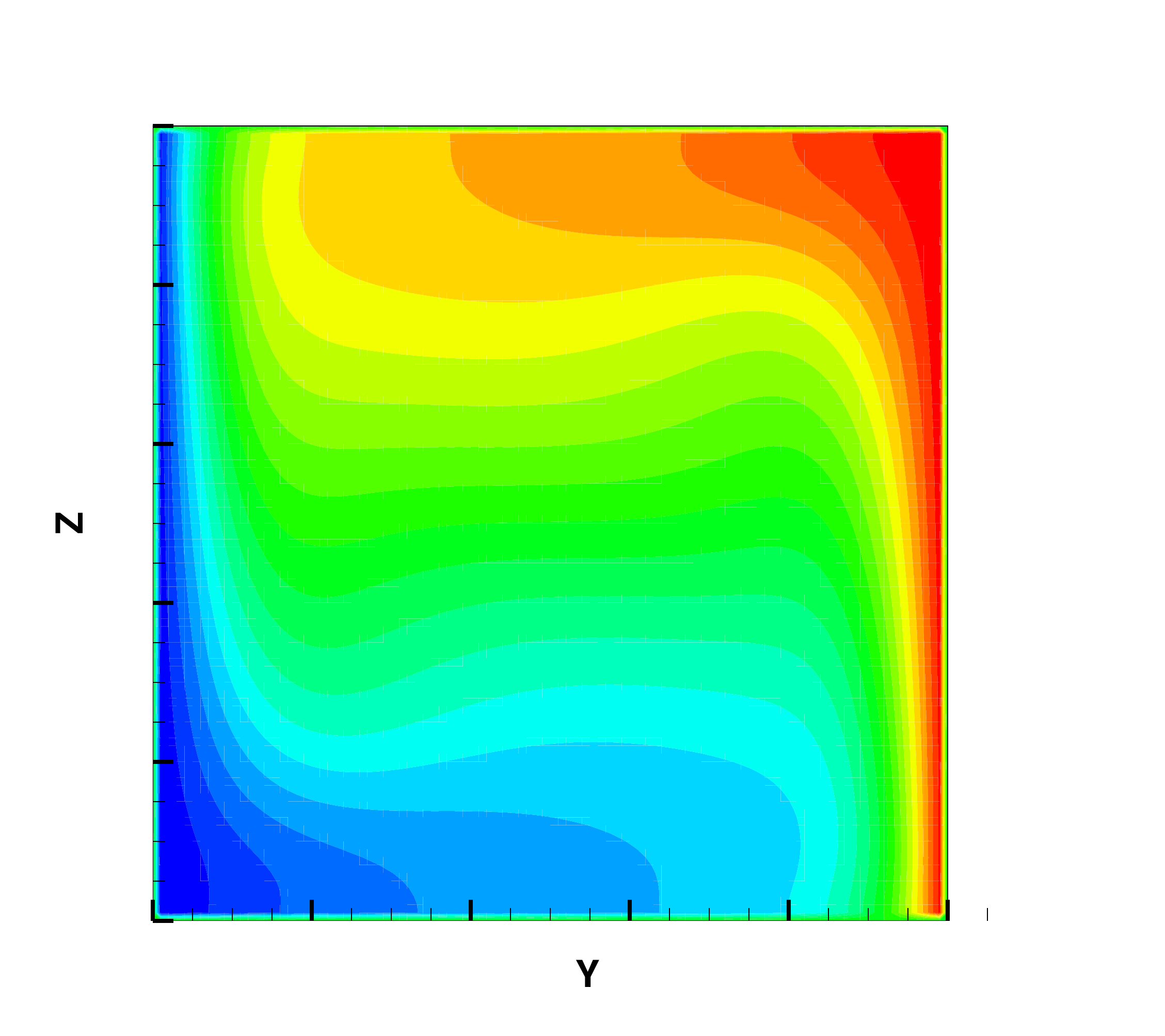}
        \label{fig:img2} } \\
%\advance\leftskip0cm
%\advance\rightskip -.62cm
             \subfloat{
        \includegraphics[width=5.5cm,scale=0.7] {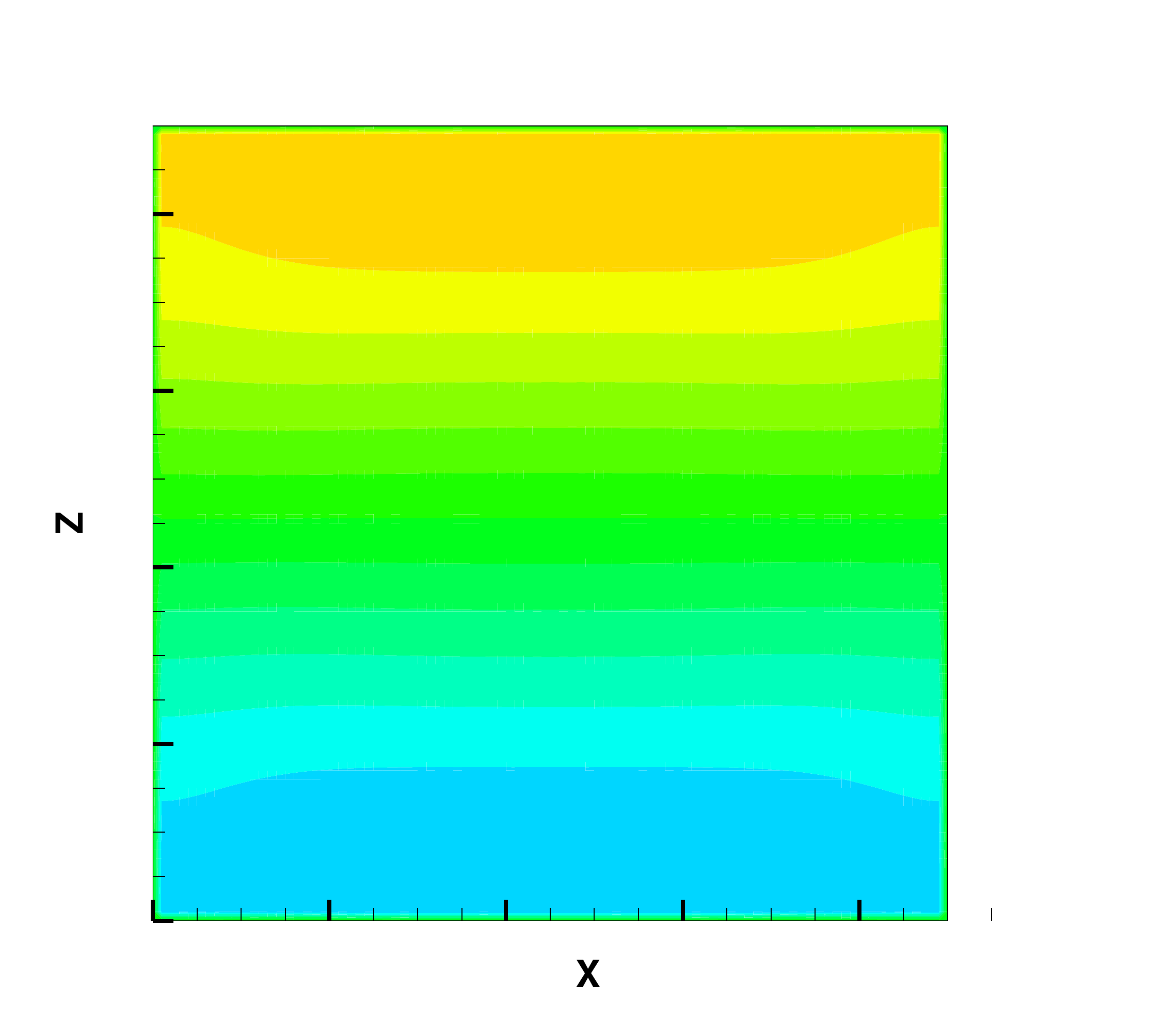}
        \label{fig:img3} } \hspace*{-35em}
     \hfill
        \subfloat{
          \includegraphics[width=5.5cm,scale=0.7] {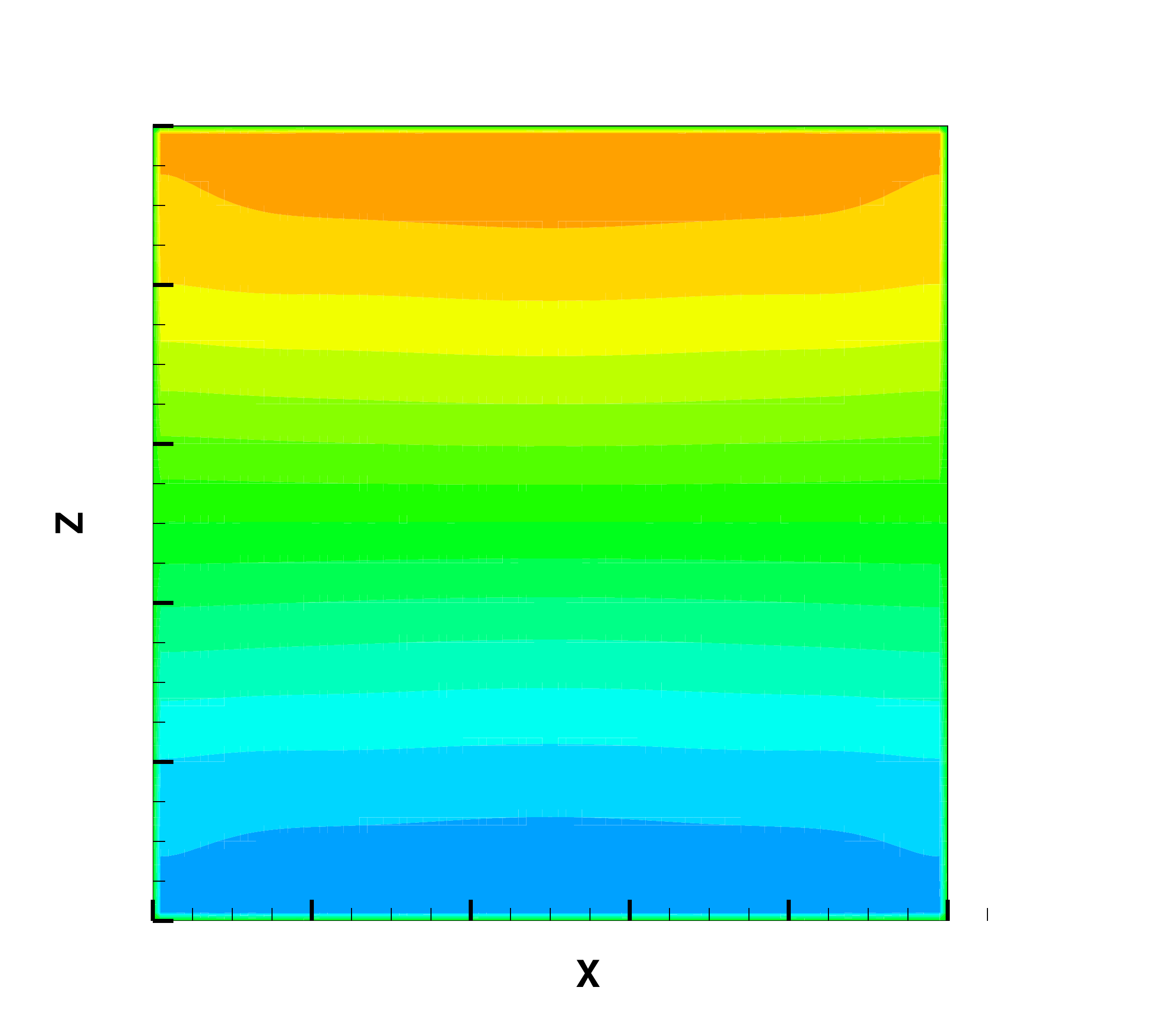}
        \label{fig:img4} }
              \\
              %\advance\leftskip-1.3cm
%\advance\rightskip -.63cm
       \subfloat{
       \includegraphics[width=5.5cm,scale=0.7]{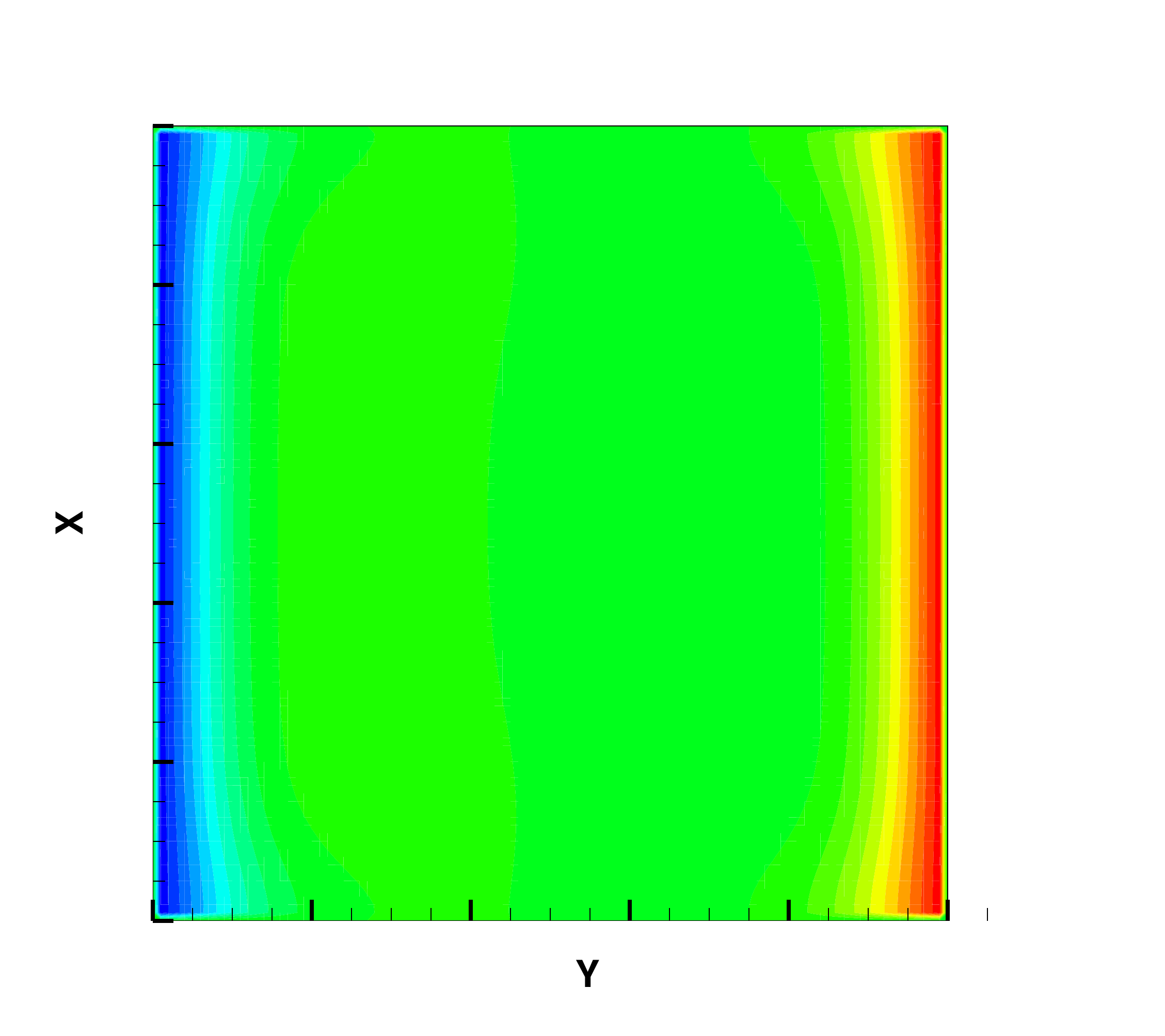}
        \label{fig:img5} }\hspace*{-35em}
             \subfloat{
         \includegraphics[width=5.5cm,scale=0.7]{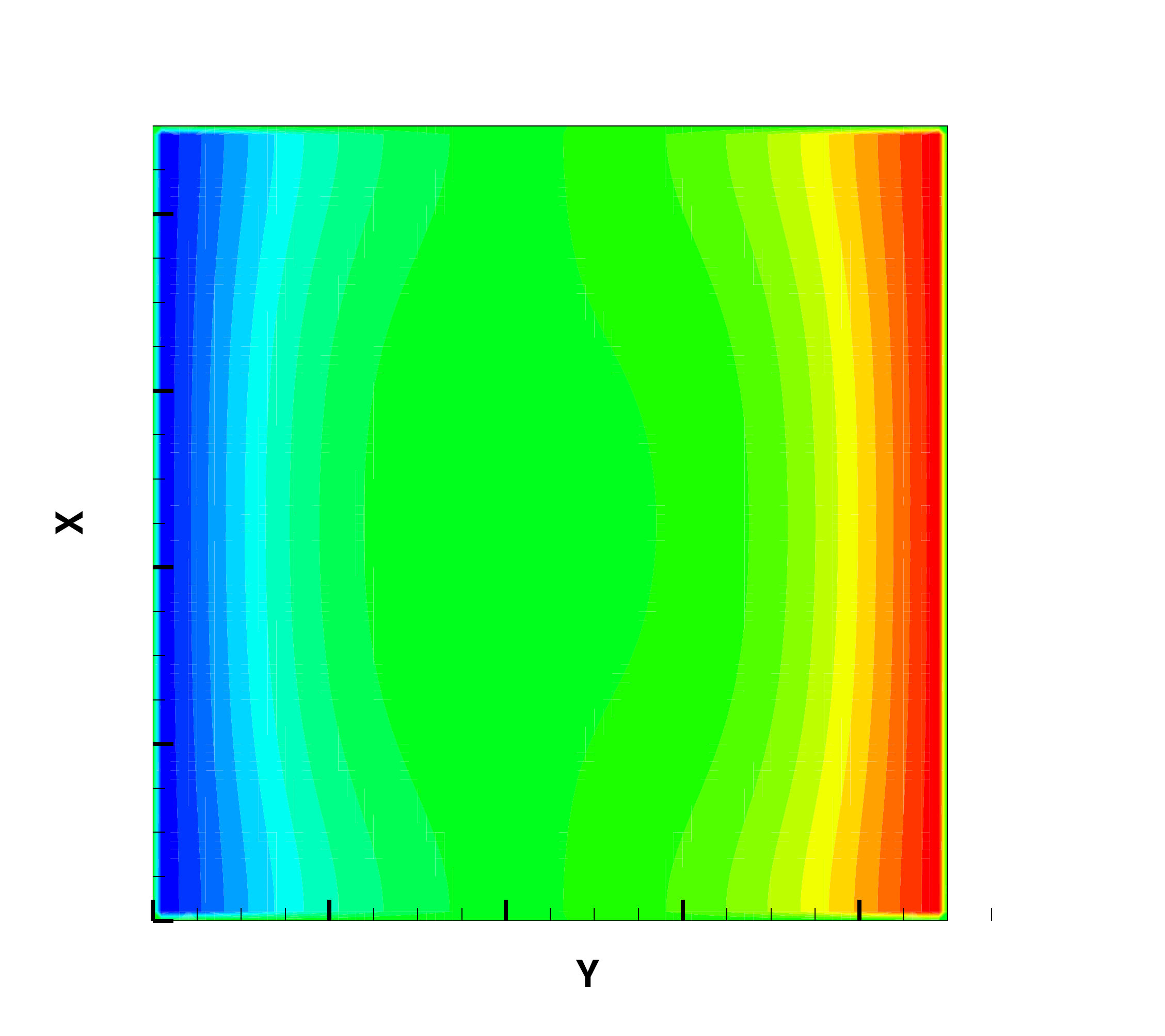}
        \label{fig:img6} }
        \caption{Temperature distribution in natural convection in a 3D cavity computed using 3D thermal cascaded LBM along different center planes at Rayleigh numbers $\mbox{Ra}=10^4$ (left) and $\mbox{Ra}=10^5$ (right). Top row: $y-z$ plane, Middle row $x-z$ plane Bottom row: $y-x$ plane.}
                \label{fig:tem}
\end{figure}
The temperature distributions, represented by isotherms, in midplanes along different directions for $\mbox{Ra}$ of $10^4$ and $10^5$ are shown in Fig.~4. It can be seen that as the natural convection effect become more significant, at higher $\mbox{Ra}=10^5$, the isotherms becomes more horizontal in the region around the center of the cavity, and becomes nearly vertical in the thin boundary layers closer to the hot and cold walls. In general, as expected, significant temperature variations appear in the thin regions in the vicinity of the isothermal wall surfaces and more uniform distributions near the adiabatic wall surfaces.

In addition, in order to provide a quantitatively study of the numerical results, we compare the following main flow and thermal characteristics of natural convection in a cubic cavity in the symmetry plane ($x=0.5$) at $\mbox{Ra}=10^3$, $10^4$ and $10^5$ computed using our 3D thermal cascaded LBM with the reference benchmark numerical solution~\cite{Fusegi1991}: The maximum horizontal velocity $u_{max}$ and its coordinate location ($y,x$), the maximum vertical velocity $w_{max}$ and its coordinate location ($y,x$); the maximum and minimum Nusselt numbers ($\mbox{Nu}_{max}$ and $\mbox{Nu}_{min}$) and their location, and, finally, the average Nusselt number $\mbox{Nu}_{mean}$. The computed results and the benchmark solutions of these quantities are presented in Table~1. It can be seen that the DDF-based 3D thermal cascaded LBM results and the benchmark solutions~\cite{Fusegi1991} are in very good quantitative agreement.
​\begin{table}[htb]
\caption{Quantitative comparison of key flow and thermal  characteristics in natural convection in a cubic cavity in the symmetry plane ($z=0.5$) between the 3D thermal cascaded LBM and the reference benchmark results obtained using a NSE solver~\cite{Fusegi1991}.}
\label{tab:nu}
\makebox[1 \textwidth][c]{
\resizebox{1.5\textwidth}{!}{\begin{tabular}{|c|cc|cc|cc|}
\hline
$\mbox{Ra}$ & \multicolumn{2}{c}{$10^3$} & \multicolumn{2}{c}{$10^4$}  &\multicolumn{2}{c|}{$10^5$} \\\hline
  $\mbox{Method}$ &3D Cascaded LBM   &     Reference Solution~\cite{Fusegi1991}       &3D Cascaded LBM             &  Reference Solution~\cite{Fusegi1991}             & 3D Cascaded LBM              & Reference Solution~\cite{Fusegi1991}          \\\hline
 $\mbox{Grid size}$  & $91\times91\times 91$ &  $31\times31\times 31$         & $91\times91\times 91$             &  $62\times62\times 62$    & $91\times91\times 91$      &  $62\times62\times 62$           \\\hline
  $u_{max}$ & 0.1308  &  0.1314 &  0.1965   & 0.2013  &   0.1441  &   0.1468  \\\hline
 $\mbox{Position}~(y,x)$  &  (0.5, 0.1910)           &    (0.5, 0.2000)        &   (0.5,0.1910)           & (0.5 , 0.1833)   &(0.5,0.1460)         &     (0.5,0.1453)         \\\hline
$w_{max}$ &  0.1308    &    0.1320   &      0.2232     &0.2252   &   0.2447  &        0.2471     \\\hline
$\mbox{Position}~(y,x)$   &   (0.8426, 0.5)  &  (0.8333, 0.5)    &  (0.8876, 0.5)     &     (0.8833, 0.5)          &     (0.9325,0.5)         &     (0.9353,0.5)          \\\hline
$\mbox{Nu}_{max}$   &      1.4170     &       1.420     &      3.5815       &  3.652           &    7.745          &    7.795       \\\hline
 $\mbox{Position}~(y,x)$  &   (0, 0.0786 )        & (0, 0.08333 )   &   ( 0,    0.1685 )        &   (0,    0.1623)            &      (0,0.0786)      &    (0,0.08256)             \\\hline
$\mbox{Nu}_{min}$    &     0.730   &     0.7639    &      0.5925        &   0.6110          & 0.772             &    0.7867     \\\hline
 $\mbox{Position}~(y,x)$  & (0, 1.0)   &    (0, 1.0)   &   (0, 1.0) &  (0, 1.0)    &      (0,1.0)    &   (0,1.0)          \\\hline
$\mbox{Nu}_{mean}$    &   1.0977   &  1.105  &  2.2647  & 2.302   &     4.4595 &  4.646 \\  \hline
  \end{tabular}}}
\end{table}

\section{Summary and Conclusions}

Fluid flows with heat transfer effects via convective transport particularly in three dimensions (3D), play  key role in a wide variety of problems of both fundamental and practical interests. Lattice Boltzmann methods are efficient computational kinetic model-based approaches that can handle such multiphysics fluid flow problems using double distribution function (DDF) formulations. In this work, we have constructed new 3D cascaded LB models using the D3Q15 and D3Q7 lattices to solve the 3D convection-diffusion based thermal energy transport equation in the DDF framework, where the fluid motion is computed from another 3D cascaded LB model from a prior study. The collision step in this approach for the transport of the temperature field is obtained by the relaxation of central moments of different orders in a multiple relaxation time formulation. This involves considering the temperature field as the only collisional invariant, while, by contrast, the LB model for the fluid flow is based on density and the components of the momentum as the conserved variables. As a result, the cascaded structure of the 3D collision operator for the solution of the temperature field is quite different from that of the flow field. In particular, the cascaded structure emerges from the second moment onwards in the former case, while for the latter case, it begins from the third order. In addition, the transport coefficient, i.e.~the thermal diffusivity for the temperature field is related to its relaxation times for the first order moments in the 3D cascaded collision model, while the kinematic viscosity of the fluid motion is dependent on the relaxation times of the second order moments of its corresponding 3D cascaded formulation. The new 3D cascaded LB models are then numerically investigated for the simulation of the 3D natural convection of air in differentially heated cubic enclosures at various Rayleigh numbers. Comparison of the computed structure of the velocity profiles and the  temperature distribution against prior numerical results show good agrement. In addition, peak convection velocities and the heat transfer rates given in terms of the Nusselt number are in good quantitative agreement with the benchmark solutions at different Rayleigh numbers.

\section*{Acknowledgements}
The authors would like to acknowledge the support of the US National Science Foundation (NSF) under Grant CBET-1705630.

\appendix
\section{Structure of the 3D Central Moments-based Collision Kernel for Fluid Flow using D3Q15 Lattice}

The details of the derivation of the 3D cascaded LBM for fluid motion with forcing terms using the D3Q15 lattice is presented in~\cite{Premnath2011three}. Here, we summarize the main results for completeness and for comparison with the corresponding 3D cascaded LB model for the solution of the transport of a scalar field represented by the CDE. The collision kernel $\widehat {\mathbf g}=(\widehat g_0,\widehat g_1,\cdots,\widehat g_{14})$ in the 3D cascaded LBE for field flow  given in Eq.~({\ref{eq:2a}}) depends on the following set of moments:

 \begin{eqnarray}
\left( {\begin{array}{*{20}{l}}
{{{\hat \kappa }_{{x^m}{y^n}{z^p}}}}^{'}\\
{\hat {\kappa} _{{x^m}{y^n}{z^p}}^{eq'}}\\
{{{\hat \sigma }_{{x^m}{y^n}{z^p}}}}^{'}\\
{{{\hat{ \bar \kappa}^{'} }_{{x^m}{y^n}{z^p}}}}
\end{array}} \right) = \sum\limits_\alpha  {\left( {\begin{array}{*{20}{l}}
{{f_\alpha }}\\
{f_\alpha ^{eq}}\\
{{S_\alpha }}\\
{{{\bar f}_\alpha }}
\end{array}} \right)} {{e^m_{\alpha x}}}{{e^n_{\alpha y}}}{{e^p_{\alpha z}}},
\label{eq:3cc}
\end{eqnarray}
where $\hat{ \bar \kappa}^{'}_{{x^m}{y^n}{z^p}}={{{\hat \kappa }_{{x^m}{y^n}{z^p}}}}^{'}-\frac {1}{2}{{{\hat \sigma }_{{x^m}{y^n}{z^p}}}}^{'}$. The specific expressions for raw moments of the source term ${{{\hat \sigma }_{{x^m}{y^n}{z^p}}}}^{'}$ as well as the corresponding source terms in the velocity space $S_\alpha$ representing the effect of a body force are presented in~\cite{Premnath2011three}. Using the notation $\hat{\bar{\eta}}^{'}_{{x^m}{y^n}{z^p}}=\hat{\bar{\kappa}}_{{x^m}{y^n}{z^p}}^{'}+\hat{\sigma}_{{x^m}{y^n}{z^p}}^{'}$ and by prescribing a central moment relaxation at different orders for the D3Q15 lattice, the structure of the collision kernel components for $\widehat {\mathbf g}$ can be expressed as (see~\cite{Premnath2011three} for details)
\begin{eqnarray}
 \widehat{g}_0&=&\widehat{g}_1=\widehat{g}_2=\widehat{g}_3=0,\\
\widehat{g}_4&=&\frac{\omega_4}{8}\left[-\widehat{\overline{\eta}}_{xy}^{'}+\rho u_xu_y+\frac{1}{2}(\widehat{\sigma}_{x}^{'}u_y+\widehat{\sigma}_{y}^{'}u_x)\right],\\
\widehat{g}_5&=&\frac{\omega_5}{8}\left[-\widehat{\overline{\eta}}_{xz}^{'}+\rho u_xu_z+\frac{1}{2}(\widehat{\sigma}_{x}^{'}u_z+\widehat{\sigma}_{z}^{'}u_x)\right],\\
\widehat{g}_6&=&\frac{\omega_6}{8}\left[-\widehat{\overline{\eta}}_{yz}^{'}+\rho u_yu_z+\frac{1}{2}(\widehat{\sigma}_{y}^{'}u_z+\widehat{\sigma}_{z}^{'}u_y)\right],\\
\widehat{g}_7&=&\frac{\omega_7}{4}\left[-(\widehat{\overline{\eta}}_{xx}^{'}-\widehat{\overline{\eta}}_{yy}^{'})+\rho (u_x^2-u_y^2) +(\widehat{\sigma}_{x}^{'}u_x-\widehat{\sigma}_{y}^{'}u_y)\right],\\
\widehat{g}_8&=&\frac{\omega_8}{12}\left[-(\widehat{\overline{\eta}}_{xx}^{'}+\widehat{\overline{\eta}}_{yy}^{'}-2\widehat{\overline{\eta}}_{zz}^{'})+\rho (u_x^2+u_y^2-2u_z^2)\right.\nonumber\\
                                      &&\left.+(\widehat{\sigma}_{x}^{'}u_x+\widehat{\sigma}_{y}^{'}u_y-2\widehat{\sigma}_{z}^{'}u_z)\right],\\
\widehat{g}_9&=&\frac{\omega_9}{18}\left[-(\widehat{\overline{\eta}}_{xx}^{'}+\widehat{\overline{\eta}}_{yy}^{'}+\widehat{\overline{\eta}}_{zz}^{'})+\rho (u_x^2+u_y^2+u_z^2)\right.\nonumber\\
                                      &&\left.+(\widehat{\sigma}_{x}^{'}u_x+\widehat{\sigma}_{y}^{'}u_y+\widehat{\sigma}_{z}^{'}u_z)+\rho\right],\\
\widehat{g}_{10}&=&\frac{\omega_{10}}{16}\left[-\widehat{\overline{\eta}}_{xyy}^{'}+2u_y\widehat{\overline{\eta}}_{xy}^{'}+u_x\widehat{\overline{\eta}}_{yy}^{'}-2\rho u_xu_y^2-\frac{1}{2}\widehat{\sigma}_{x}^{'}u_y^2-\widehat{\sigma}_{y}^{'}u_yu_x\right]\nonumber\\
                                      &&+u_y\widehat{g}_4+\frac{1}{8}u_x(-\widehat{g}_7+\widehat{g}_8+3\widehat{g}_9),\allowdisplaybreaks\\
\widehat{g}_{11}&=&\frac{\omega_{11}}{16}\left[-\widehat{\overline{\eta}}_{xxy}^{'}+2u_x\widehat{\overline{\eta}}_{xy}^{'}+u_y\widehat{\overline{\eta}}_{xx}^{'}-2\rho u_x^2u_y-\frac{1}{2}\widehat{\sigma}_{y}^{'}u_x^2-\widehat{\sigma}_{x}^{'}u_xu_y\right]\nonumber\\
                                      &&+u_x\widehat{g}_4+\frac{1}{8}u_y(\widehat{g}_7+\widehat{g}_8+3\widehat{g}_9),\\
\widehat{g}_{12}&=&\frac{\omega_{12}}{16}\left[-\widehat{\overline{\eta}}_{xxz}^{'}+2u_x\widehat{\overline{\eta}}_{xz}^{'}+u_z\widehat{\overline{\eta}}_{xx}^{'}-2\rho u_x^2u_z-\frac{1}{2}\widehat{\sigma}_{z}^{'}u_x^2-\widehat{\sigma}_{x}^{'}u_xu_z\right]\nonumber\\
                                      &&+u_x\widehat{g}_5+\frac{1}{8}u_z(\widehat{g}_7+\widehat{g}_8+3\widehat{g}_9), \allowdisplaybreaks\\
\widehat{g}_{13}&=&\frac{\omega_{13}}{8}\left[-\widehat{\overline{\eta}}_{xyz}^{'}+u_x\widehat{\overline{\eta}}_{yz}^{'}+u_y\widehat{\overline{\eta}}_{xz}^{'}+u_z\widehat{\overline{\eta}}_{xy}^{'}-2\rho u_xu_yu_z-\frac{1}{2}\left(\widehat{\sigma}_{x}^{'}u_yu_z\right.\right.\nonumber\\
                 &&\left.\left.+\widehat{\sigma}_{y}^{'}u_xu_z+\widehat{\sigma}_{z}^{'}u_xu_y\right)\right]
                    +u_z\widehat{g}_4+u_y\widehat{g}_5+u_x\widehat{g}_6,\\
\widehat{g}_{14}&=&\frac{\omega_{14}}{16}\left[-\widehat{\overline{\eta}}_{xxyy}^{'}+2u_x\widehat{\overline{\eta}}_{xyy}^{'}+2u_y\widehat{\overline{\eta}}_{xxy}^{'}
                    -u_x^2\widehat{\overline{\eta}}_{yy}^{'}-u_y^2\widehat{\overline{\eta}}_{xx}^{'}-4u_xu_y\widehat{\overline{\eta}}_{xy}^{'}\right.\nonumber\\
                    &&\left.+\widetilde{\widehat{\kappa}}_{xx}\widetilde{\widehat{\kappa}}_{yy}+3\rho u_x^2u_y^2+\widehat{\sigma}_{x}^{'}u_xu_y^2+\widehat{\sigma}_{y}^{'}u_yu_x^2\right]-2u_xu_y\widehat{g}_{4}+\frac{1}{8}(u_x^2-u_y^2)\widehat{g}_{7}\nonumber\\
                    &&+\frac{1}{8}(-u_x^2-u_y^2)\widehat{g}_{8}+\left(\frac{3}{8}(-u_x^2-u_y^2)-\frac{1}{2}\right)\widehat{g}_{9}+2u_x\widehat{g}_{10}+2u_y\widehat{g}_{11},
\end{eqnarray}
where $\omega_4,\omega_5,\ldots,\omega_{14}$  are the relaxation parameters~($0<\omega_\beta<2 $ for $\beta=4,5,\cdots,14$). The relaxation times for the second order moments are related to the kinematic viscosity $\nu$ of the fluid being simulated through $\nu=c_s^2(\frac{1}{\omega_j}-\frac{1}{2})$, where $j=5,\cdots,9$. In this work, we choose $c_s^2=1/3$. The rest of the relaxation parameters, which influence numerical stability, are set to unity in the present work. Finally, by expanding the product $(\tensor K \cdot \widehat{\mathbf g})_\alpha$ in Eq.~({\ref{eq:2a}}), the post-collision values of the distribution function are given by
 \begin{eqnarray}
\widetilde{\overline{f}}_{0}&=&\overline{f}_{0}+\left[\widehat{g}_0-2\widehat{g}_9+32\widehat{g}_{14}\right]+S_0, \nonumber\\
\widetilde{\overline{f}}_{1}&=&\overline{f}_{1}+\left[\widehat{g}_0+\widehat{g}_1+\widehat{g}_{7}+\widehat{g}_{8}-\widehat{g}_{9}-8\widehat{g}_{10}-8\widehat{g}_{14}\right]+S_1,\nonumber\\
\widetilde{\overline{f}}_{2}&=&\overline{f}_{2}+\left[\widehat{g}_0-\widehat{g}_1+\widehat{g}_{7}+\widehat{g}_{8}-\widehat{g}_{9}+8\widehat{g}_{10}-8\widehat{g}_{14}\right]+S_2,\nonumber\\
\widetilde{\overline{f}}_{3}&=&\overline{f}_{3}+\left[\widehat{g}_0+\widehat{g}_2-\widehat{g}_{7}+\widehat{g}_{8}-\widehat{g}_{9}-8\widehat{g}_{11}-8\widehat{g}_{14}\right]+S_3,\nonumber\\
\widetilde{\overline{f}}_{4}&=&\overline{f}_{4}+\left[\widehat{g}_0-\widehat{g}_2-\widehat{g}_{7}+\widehat{g}_{8}-\widehat{g}_{9}+8\widehat{g}_{11}-8\widehat{g}_{14}\right]+S_4,\nonumber\\
\widetilde{\overline{f}}_{5}&=&\overline{f}_{5}+\left[\widehat{g}_0+\widehat{g}_3-2\widehat{g}_{8}-\widehat{g}_{9}-8\widehat{g}_{12}-8\widehat{g}_{14}\right]+S_5,\nonumber\\
\widetilde{\overline{f}}_{6}&=&\overline{f}_{6}+\left[\widehat{g}_0-\widehat{g}_3-2\widehat{g}_{8}-\widehat{g}_{9}+8\widehat{g}_{12}-8\widehat{g}_{14}\right]+S_6,\nonumber\\
\widetilde{\overline{f}}_{7}&=&\overline{f}_{7}+\left[\widehat{g}_0+\widehat{g}_1+\widehat{g}_2+\widehat{g}_3+\widehat{g}_4+\widehat{g}_5+\widehat{g}_6+\widehat{g}_9+2\widehat{g}_{10}+2\widehat{g}_{11}+2\widehat{g}_{12}\right.\nonumber\\
                                &&\left.+\widehat{g}_{13}+2\widehat{g}_{14}\right]+S_7,\nonumber\\
\widetilde{\overline{f}}_{8}&=&\overline{f}_{8}+\left[\widehat{g}_0-\widehat{g}_1+\widehat{g}_2+\widehat{g}_3-\widehat{g}_4-\widehat{g}_5+\widehat{g}_6+\widehat{g}_9-2\widehat{g}_{10}+2\widehat{g}_{11}+2\widehat{g}_{12}\right.\nonumber\\
                                &&\left.-\widehat{g}_{13}+2\widehat{g}_{14}\right]+S_8,\nonumber\\
\widetilde{\overline{f}}_{9}&=&\overline{f}_{9}+\left[\widehat{g}_0+\widehat{g}_1-\widehat{g}_2+\widehat{g}_3-\widehat{g}_4+\widehat{g}_5-\widehat{g}_6+\widehat{g}_9+2\widehat{g}_{10}-2\widehat{g}_{11}+2\widehat{g}_{12}\right.\nonumber\\
                                &&\left.-\widehat{g}_{13}+2\widehat{g}_{14}\right]+S_9,\nonumber \allowdisplaybreaks \\
\widetilde{\overline{f}}_{10}&=&\overline{f}_{10}+\left[\widehat{g}_0-\widehat{g}_1-\widehat{g}_2+\widehat{g}_3+\widehat{g}_4-\widehat{g}_5-\widehat{g}_6+\widehat{g}_9-2\widehat{g}_{10}-2\widehat{g}_{11}+2\widehat{g}_{12}\right.\nonumber\\
                                &&\left.+\widehat{g}_{13}+2\widehat{g}_{14}\right]+S_{10},\nonumber\\
\widetilde{\overline{f}}_{11}&=&\overline{f}_{11}+\left[\widehat{g}_0+\widehat{g}_1+\widehat{g}_2-\widehat{g}_3+\widehat{g}_4-\widehat{g}_5-\widehat{g}_6+\widehat{g}_9+2\widehat{g}_{10}+2\widehat{g}_{11}-2\widehat{g}_{12}\right.\nonumber\\
                                &&\left.-\widehat{g}_{13}+2\widehat{g}_{14}\right]+S_{11},\nonumber\\
\widetilde{\overline{f}}_{12}&=&\overline{f}_{12}+\left[\widehat{g}_0-\widehat{g}_1+\widehat{g}_2-\widehat{g}_3-\widehat{g}_4+\widehat{g}_5-\widehat{g}_6+\widehat{g}_9-2\widehat{g}_{10}+2\widehat{g}_{11}-2\widehat{g}_{12}\right.\nonumber\\
                                &&\left.-\widehat{g}_{13}+2\widehat{g}_{14}\right]+S_{12},\nonumber\\
\widetilde{\overline{f}}_{13}&=&\overline{f}_{13}+\left[\widehat{g}_0+\widehat{g}_1-\widehat{g}_2-\widehat{g}_3-\widehat{g}_4-\widehat{g}_5+\widehat{g}_6+\widehat{g}_9+2\widehat{g}_{10}-2\widehat{g}_{11}-2\widehat{g}_{12}\right.\nonumber\\
                                &&\left.+\widehat{g}_{13}+2\widehat{g}_{14}\right]+S_{13},\nonumber\allowdisplaybreaks\\
\widetilde{\overline{f}}_{14}&=&\overline{f}_{14}+\left[\widehat{g}_0-\widehat{g}_1-\widehat{g}_2-\widehat{g}_3+\widehat{g}_4+\widehat{g}_5+\widehat{g}_6+\widehat{g}_9-2\widehat{g}_{10}-2\widehat{g}_{11}-2\widehat{g}_{12}\right.\nonumber\\
                                &&\left.-\widehat{g}_{13}+2\widehat{g}_{14}\right]+S_{14}.
\end{eqnarray}
Then, after performing the streaming step as given in Eq.~({\ref{eq:2b}}), we get the updated distribution function from which the velocity field $\bm u$ can be computed as shown in Eq.~({\ref{eq:3}}).

\section{Source Terms for the 3D Cascaded LBE for Scalar Field using D3Q15 Lattice}
Using the source moments projected to the orthogonal basis vectors $\widehat {\mathbf{m}}^{s,\phi}$ defined in Eq.~(\ref{eq:19}) and inverting it by using  $\mathbf S^\phi=\tensor K^{-1}\cdot \widehat{\mathbf m}^{s,\phi}$, and exploiting the orthogonality of the collision matrix $\tensor K$, we get following expressions for the source terms in the velocity space for the D3Q15 lattice used in the solution of the 3D CDE:
\begin{eqnarray}
S_0^\phi&=&\frac{1}{45}\left[3\widehat{m}^{s,\phi}_{0}-5\widehat{m}^{s,\phi}_{9}+\widehat{m}^{s,\phi}_{14}\right], \nonumber\\
S_1^\phi&=&\frac{1}{180}\left[12\widehat{m}^{s,\phi}_{0}+18\widehat{m}^{s,\phi}_{1}+45\widehat{m}^{s,\phi}_{7}+15\widehat{m}^{s,\phi}_{8}-10\widehat{m}^{s,\phi}_{9}-9\widehat{m}^{s,\phi}_{10}-\widehat{m}^{s,\phi}_{14}\right],\nonumber\\
S_2^\phi&=&\frac{1}{180}\left[12\widehat{m}^{s,\phi}_{0}-18\widehat{m}^{s,\phi}_{1}+45\widehat{m}^{s,\phi}_{7}+15\widehat{m}^{s,\phi}_{8}-10\widehat{m}^{s,\phi}_{9}+9\widehat{m}^{s,\phi}_{10}-\widehat{m}^{s,\phi}_{14}\right],\nonumber\\
S_3^\phi&=&\frac{1}{180}\left[12\widehat{m}^{s,\phi}_{0}+18\widehat{m}^{s,\phi}_{2}-45\widehat{m}^{s,\phi}_{7}+15\widehat{m}^{s,\phi}_{8}-10\widehat{m}^{s,\phi}_{9}-9\widehat{m}^{s,\phi}_{11}-\widehat{m}^{s,\phi}_{14}\right],\nonumber\\
S_4^\phi&=&\frac{1}{180}\left[12\widehat{m}^{s,\phi}_{0}-18\widehat{m}^{s,\phi}_{2}-45\widehat{m}^{s,\phi}_{7}+15\widehat{m}^{s,\phi}_{8}-10\widehat{m}^{s,\phi}_{9}+9\widehat{m}^{s,\phi}_{11}-\widehat{m}^{s,\phi}_{14}\right],\nonumber\\
S_5^\phi&=&\frac{1}{180}\left[12\widehat{m}^{s,\phi}_{0}+18\widehat{m}^{s,\phi}_{3}-30\widehat{m}^{s,\phi}_{8}-10\widehat{m}^{s,\phi}_{9}-9\widehat{m}^{s,\phi}_{12}-\widehat{m}^{s,\phi}_{14}\right],\nonumber\\
S_6^\phi&=&\frac{1}{180}\left[12\widehat{m}^{s,\phi}_{0}-18\widehat{m}^{s,\phi}_{3}-30\widehat{m}^{s,\phi}_{8}-10\widehat{m}^{s,\phi}_{9}+9\widehat{m}^{s,\phi}_{12}-\widehat{m}^{s,\phi}_{14}\right],\nonumber\\
S_7^\phi&=&\frac{1}{720}\left[48\widehat{m}^{s,\phi}_{0}+72\widehat{m}^{s,\phi}_{1}+72\widehat{m}^{s,\phi}_{2}+72\widehat{m}^{s,\phi}_{3}+90\widehat{m}^{s,\phi}_{4}+90\widehat{m}^{s,\phi}_{5}+90\widehat{m}^{s,\phi}_{6}+40\widehat{m}^{s,\phi}_{9}\right.\nonumber\\
                 &&\left.+9\widehat{m}^{s,\phi}_{10}+9\widehat{m}^{s,\phi}_{11}+9\widehat{m}^{s,\phi}_{12}+90\widehat{m}^{s,\phi}_{13}+\widehat{m}^{s,\phi}_{14}\right],\nonumber\\
S_8^\phi&=&\frac{1}{720}\left[48\widehat{m}^{s,\phi}_{0}-72\widehat{m}^{s,\phi}_{1}+72\widehat{m}^{s,\phi}_{2}+72\widehat{m}^{s,\phi}_{3}-90\widehat{m}^{s,\phi}_{4}-90\widehat{m}^{s,\phi}_{5}+90\widehat{m}^{s,\phi}_{6}+40\widehat{m}^{s,\phi}_{9}\right.\nonumber\\
                 &&\left.-9\widehat{m}^{s,\phi}_{10}+9\widehat{m}^{s,\phi}_{11}+9\widehat{m}^{s,\phi}_{12}-90\widehat{m}^{s,\phi}_{13}+\widehat{m}^{s,\phi}_{14}\right],\nonumber \\
S_9^\phi&=&\frac{1}{720}\left[48\widehat{m}^{s,\phi}_{0}+72\widehat{m}^{s,\phi}_{1}-72\widehat{m}^{s,\phi}_{2}+72\widehat{m}^{s,\phi}_{3}-90\widehat{m}^{s,\phi}_{4}+90\widehat{m}^{s,\phi}_{5}-90\widehat{m}^{s,\phi}_{6}+40\widehat{m}^{s,\phi}_{9}\right.\nonumber\\
                 &&\left.+9\widehat{m}^{s,\phi}_{10}-9\widehat{m}^{s,\phi}_{11}+9\widehat{m}^{s,\phi}_{12}-90\widehat{m}^{s,\phi}_{13}+\widehat{m}^{s,\phi}_{14}\right],\nonumber\\
S_{10}^\phi&=&\frac{1}{720}\left[48\widehat{m}^{s,\phi}_{0}-72\widehat{m}^{s,\phi}_{1}-72\widehat{m}^{s,\phi}_{2}+72\widehat{m}^{s,\phi}_{3}+90\widehat{m}^{s,\phi}_{4}-90\widehat{m}^{s,\phi}_{5}-90\widehat{m}^{s,\phi}_{6}+40\widehat{m}^{s,\phi}_{9}\right.\nonumber\\
                 &&\left.-9\widehat{m}^{s,\phi}_{10}-9\widehat{m}^{s,\phi}_{11}+9\widehat{m}^{s,\phi}_{12}+90\widehat{m}^{s,\phi}_{13}+\widehat{m}^{s,\phi}_{14}\right],\nonumber \allowdisplaybreaks\\
S_{11}^\phi&=&\frac{1}{720}\left[48\widehat{m}^{s,\phi}_{0}+72\widehat{m}^{s,\phi}_{1}+72\widehat{m}^{s,\phi}_{2}-72\widehat{m}^{s,\phi}_{3}+90\widehat{m}^{s,\phi}_{4}-90\widehat{m}^{s,\phi}_{5}-90\widehat{m}^{s,\phi}_{6}+40\widehat{m}^{s,\phi}_{9}\right.\nonumber\\
                 &&\left.+9\widehat{m}^{s,\phi}_{10}+9\widehat{m}^{s,\phi}_{11}-9\widehat{m}^{s,\phi}_{12}-90\widehat{m}^{s,\phi}_{13}+\widehat{m}^{s,\phi}_{14}\right],\nonumber\\
S_{12}^\phi&=&\frac{1}{720}\left[48\widehat{m}^{s,\phi}_{0}-72\widehat{m}^{s,\phi}_{1}+72\widehat{m}^{s,\phi}_{2}-72\widehat{m}^{s,\phi}_{3}-90\widehat{m}^{s,\phi}_{4}+90\widehat{m}^{s,\phi}_{5}-90\widehat{m}^{s,\phi}_{6}+40\widehat{m}^{s,\phi}_{9}\right.\nonumber\\
                 &&\left.-9\widehat{m}^{s,\phi}_{10}+9\widehat{m}^{s,\phi}_{11}-9\widehat{m}^{s,\phi}_{12}+90\widehat{m}^{s,\phi}_{13}+\widehat{m}^{s,\phi}_{14}\right],\nonumber\\
S_{13}^\phi&=&\frac{1}{720}\left[48\widehat{m}^{s,\phi}_{0}+72\widehat{m}^{s,\phi}_{1}-72\widehat{m}^{s,\phi}_{2}-72\widehat{m}^{s,\phi}_{3}-90\widehat{m}^{s,\phi}_{4}-90\widehat{m}^{s,\phi}_{5}+90\widehat{m}^{s,\phi}_{6}+40\widehat{m}^{s,\phi}_{9}\right.\nonumber\\
                 &&\left.+9\widehat{m}^{s,\phi}_{10}-9\widehat{m}^{s,\phi}_{11}-9\widehat{m}^{s,\phi}_{12}+90\widehat{m}^{s,\phi}_{13}+\widehat{m}^{s,\phi}_{14}\right],\nonumber\\
S_{14}^\phi&=&\frac{1}{720}\left[48\widehat{m}^{s,\phi}_{0}-72\widehat{m}^{s,\phi}_{1}-72\widehat{m}^{s,\phi}_{2}-72\widehat{m}^{s,\phi}_{3}+90\widehat{m}^{s,\phi}_{4}+90\widehat{m}^{s,\phi}_{5}+90\widehat{m}^{s,\phi}_{6}+40\widehat{m}^{s,\phi}_{9}\right.\nonumber\\
                 &&\left.-9\widehat{m}^{s,\phi}_{10}-9\widehat{m}^{s,\phi}_{11}-9\widehat{m}^{s,\phi}_{12}-90\widehat{m}^{s,\phi}_{13}+\widehat{m}^{s,\phi}_{14}\right].
\end{eqnarray}

\section{3D Cascaded LB Model for Transport of Temperature Field using D3Q7 Lattice}
The CDE for the scalar field $\phi$, such as the temperature, given in Eq.~({\ref{eq:4}}) has the diffusion term with lower degree of symmetry than that of the viscous stress tensor term in the NSE. As a result, the lattice set to represent the CDE can possibly satisfy lower degree of symmetry and isotropy requirements than that for the NSE. Hence, one can  also construct a simplified 3D cascaded LBE for the CDE using a three-dimensional, seven velocity (D3Q7) lattice. In this regard, the components of the particle velocity along with the unit vector for this lattice are given by
\begin{eqnarray}
&\ket{e_{\alpha x}} =\left(     0,     1,    -1,     0,     0,  0,0 \right)^\dag, \nonumber\\
&\ket{e_{\alpha y}} =\left(     0,     0,    0,     1,     -1,  0,0 \right)^\dag,
\nonumber\\
&\ket{e_{\alpha z}} =\left(     0,     0,    0,     0,     0,  1,-1 \right)^\dag,
\nonumber\\
&\ket{\phi} =\left(     1,     1,    1,     1,     1,  1,1 \right)^\dag.
\label{eq:a1}
\end{eqnarray}
Starting from the nominal basis vectors
\begin{eqnarray*}
&\mathbf T_{0}=\ket{\phi},\\
&\mathbf T_{1}=\ket{e_{\alpha x}},\quad
\mathbf T_{2}=\ket{e_{\alpha y}},\quad
\mathbf T_{3}=\ket{e_{\alpha z}},\nonumber\\
&\mathbf T_{4}=\ket{e_{\alpha x}^2-e_{\alpha y}^2},\quad
\mathbf T_{5}=\ket{e_{\alpha x}^2-e_{\alpha z}^2}\\
&\mathbf T_{6}=\ket{e_{\alpha x}^2+e_{\alpha y}^2+e_{\alpha z}^2},
\label{eq:c2}
\end{eqnarray*}
and applying the Gram-Schmidt procedure, the corresponding linearly independent orthogonal basis vectors are given by
\begin{eqnarray*}
&\mathbf K_{0}=\ket{\phi}\\
&\mathbf K_{1}=\ket{e_{\alpha x}},\quad
\mathbf K_{2}=\ket{e_{\alpha y}},\quad
\mathbf K_{3}=\ket{e_{\alpha z}},\quad
\mathbf K_{4}=\ket{e_{\alpha x}^2-e_{\alpha y}^2},\\
&\mathbf K_{5}=2\ket{e_{\alpha x}^2-e_{\alpha z}^2}-\ket{e_{\alpha x}^2-e_{\alpha y}^2},\quad
\mathbf K_{6}=7\ket{(e_{\alpha x}^2+e_{\alpha y}^2+e_{\alpha z}^2)}-6\ket{\phi}.
\label{eq:c3}
\end{eqnarray*}
Next, the orthogonal collision matrix can be written as
\begin{eqnarray}
\tensor{K}&=&\left[{{\mathbf{K}}_{0}},{\mathbf{K}_{1}},{\mathbf{K}_{2}},{\mathbf{K}_{3}},{\mathbf{K}_{4}},{\mathbf{K}_{5}},{\mathbf{K}_{6}},{\mathbf{K}_{7}}      \right].
\label{eq:c4}
\end{eqnarray}
The discrete central moments of various quantities and their corresponding raw moments are given in Eq.~(\ref{eq:14}) and Eq.~(\ref{eq:17}), respectively, where $\alpha=0,1,\cdots,6$ is considered. Following the overall procedure discussed in Sec.~2 and adopting it for the D3Q7 lattice, various results can now be summarized. First, the raw moments of the source term at different orders are given by
\begin{eqnarray}
&\hat {\sigma} _{0}^{\phi'}=R,\nonumber \\
&\hat {\sigma}_{x}^{\phi'}=u_xR,\quad \hat {\sigma} _{y}^{\phi'}=u_y R,\quad \hat {\sigma} _{z}^{\phi'}=u_z R,\nonumber \\
&\hat {\sigma}_{xx}^{\phi'}=u_x^2R,\quad \hat {\sigma} _{yy}^{\phi'}=u_y^2 R,\quad \hat {\sigma} _{zz}^{\phi'}=u_z^2 R.
\end{eqnarray}
Then, as in Sec.~2 transforming them to the velocity space, the source term in the particle velocity space are given as
\begin{eqnarray}
S_0^\phi&=&\frac{1}{7}\left[\widehat{m}^{s,\phi}_{0}-\widehat{m}^{s,\phi}_{6}\right], \nonumber\\
S_1^\phi&=&\frac{1}{84}\left[12\widehat{m}^{s,\phi}_{0}+42\widehat{m}^{s,\phi}_{1}+21\widehat{m}^{s,\phi}_{4}+7\widehat{m}^{s,\phi}_{5}+2\widehat{m}^{s,\phi}_{6}\right],\nonumber\\
S_2^\phi&=&\frac{1}{84}\left[12\widehat{m}^{s,\phi}_{0}-42\widehat{m}^{s,\phi}_{1}+21\widehat{m}^{s,\phi}_{4}+7\widehat{m}^{s,\phi}_{5}+2\widehat{m}^{s,\phi}_{6}\right],\nonumber\\
S_3^\phi&=&\frac{1}{84}\left[12\widehat{m}^{s,\phi}_{0}+42\widehat{m}^{s,\phi}_{1}-21\widehat{m}^{s,\phi}_{4}+7\widehat{m}^{s,\phi}_{5}+2\widehat{m}^{s,\phi}_{6}\right],\nonumber\\
S_4^\phi&=&\frac{1}{84}\left[12\widehat{m}^{s,\phi}_{0}-42\widehat{m}^{s,\phi}_{1}-21\widehat{m}^{s,\phi}_{4}+7\widehat{m}^{s,\phi}_{5}+2\widehat{m}^{s,\phi}_{6}\right],\nonumber\\
S_5^\phi&=&\frac{1}{42}\left[6\widehat{m}^{s,\phi}_{0}+21\widehat{m}^{s,\phi}_{3}-7\widehat{m}^{s,\phi}_{5}+\widehat{m}^{s,\phi}_{6}\right],\nonumber\\
S_6^\phi&=&\frac{1}{42}\left[6\widehat{m}^{s,\phi}_{0}-21\widehat{m}^{s,\phi}_{3}-7\widehat{m}^{s,\phi}_{5}+\widehat{m}^{s,\phi}_{6}\right],
\end{eqnarray}
where
\begin{eqnarray}
&\widehat{m}^{s,\phi}_{0}=\braket{K_0|S_{\alpha}^\phi}=R, \quad
\widehat{m}^{s,\phi}_{1}=\braket{K_1|S_{\alpha}^\phi}=u_xR, \quad
\widehat{m}^{s,\phi}_{2}=\braket{K_2|S_{\alpha}^\phi}=u_yR, \nonumber\\
&\widehat{m}^{s,\phi}_{3}=\braket{K_3|S_{\alpha}^\phi}=u_zR, \quad
\widehat{m}^{s,\phi}_{4}=\braket{K_4|S_{\alpha}^\phi}=(u_x^2-u_y^2)R, \nonumber\\
&\widehat{m}^{s,\phi}_{5}=\braket{K_5|S_{\alpha}^\phi}=(u_x^2+u_y^2-2 uz^2)R , \nonumber\\
&\widehat{m}^{s,\phi}_{6}=\braket{K_6|S_{\alpha}^\phi}=(7u_x^2+u_y^2+u_z^2-6)R.
\end{eqnarray}
Then, by prescribing the relaxation of central moments to their corresponding equilibria for first and higher orders, and following the approach presented in Sec.~2, we get the collision kernel for the D3Q7 lattice as
\begin{eqnarray}
\widehat{h}_1&=&0,\nonumber\\
\widehat{h}_1&=&\frac{\omega_1^\phi}{2}\left[\phi u_x-\widehat{\overline{\kappa}}_{x}^{\phi'}- u_x (R/2) \right],\nonumber\\
\widehat{h}_2&=&\frac{\omega_2^\phi}{2}\left[\phi u_y-\widehat{\overline{\kappa}}_{y}^{\phi'}- u_y (R/2) \right],\nonumber\\
\widehat{h}_3&=&\frac{\omega_3^\phi}{2}\left[\phi u_z-\widehat{\overline{\kappa}}_{z}^{\phi'}- u_z (R/2) \right],\nonumber\\
\widehat{h}_4&=&\frac{\omega_4^\phi}{4}\left[-(\widehat{\overline{\eta}}_{xx}^{\phi'}-\widehat{\overline{\eta}}_{yy}^{\phi'})+2(u_x\widehat{\overline{\eta}}_{x}^{\phi'}-u_y\widehat{\overline{\eta}}_{y}^{\phi'})-\left(\phi+\frac{R}{2}\right)(u_x^2-u_y^2)\right]+ \nonumber\\
                                      &&u_x\widehat{h}_1-u_y\widehat{h}_2 ,\nonumber\\
\widehat{h}_5&=&\frac{\omega_5^\phi}{12}\left[-(\widehat{\overline{\eta}}_{xx}^{\phi'}+\widehat{\overline{\eta}}_{yy}^{\phi'}-2\widehat{\overline{\eta}}_{zz}^{\phi'})+2(u_x\widehat{\overline{\eta}}_{x}^{\phi'}+u_y\widehat{\overline{\eta}}_{y}^{\phi'}-2u_z\widehat{\overline{\eta}}_{z}^{\phi'})-\left(\phi+\frac{R}{2}\right)(u_x^2+u_y^2-2u_z^2)\right]+ \nonumber\\
                                      &&\frac{1}{3}\left(u_x\widehat{h}_1+u_y\widehat{h}_2-2u_z\widehat{h}_3\right),\nonumber \\
\widehat{h}_6&=&\frac{\omega_6^\phi}{6}\left[3c_{s\phi}^2\phi-(\widehat{\overline{\eta}}_{xx}^{\phi'}+\widehat{\overline{\eta}}_{yy}^{\phi'}+\widehat{\overline{\eta}}_{zz}^{\phi'})+2(u_x\widehat{\overline{\eta}}_{x}^{\phi'}+u_y\widehat{\overline{\eta}}_{y}^{\phi'}+u_z\widehat{\overline{\eta}}_{z}^{\phi'})-\left(\phi+\frac{R}{2}\right)(u_x^2+u_y^2+u_z^2)\right]+ \nonumber\\
                                      &&\frac{2}{3}\left(u_x\widehat{h}_1+u_y\widehat{h}_2+u_z\widehat{h}_3\right),
\end{eqnarray}
where $\omega_1^\phi,\omega_2^\phi,\ldots,\omega_6^\phi$ are relaxation parameters and the definition given in Eq.~(\ref{eq:21}) for the raw moment
$\widehat{\overline{\eta}}_{x^my^nz^p}^{\phi'}$ is used for compact representation. The coefficient of diffusivity of the scalar field in the CDE, i.e. $D_\phi$ in Eq.~(\ref{eq:4}) is related to the relaxation parameters for the first order moments through $D_\phi=c_{s\phi}^2(\frac{1}{\omega_j^\phi}-\frac{1}{2})$, where $j=1,2,3$. Finally, the post-collision values of the distribution function are obtained from Eq.~(\ref{eq:13a}) after expanding $(\tensor K \cdot \widehat{\mathbf h})_\alpha$ for the D3Q7 lattice as
\begin{eqnarray}
\widetilde{\overline{g}}_{0}&=&\overline{g}_{0}+\left[\widehat{h}_0-6\widehat{h}_6\right]+S^\phi_0, \nonumber\\
\widetilde{\overline{g}}_{1}&=&\overline{g}_{1}+\left[\widehat{h}_0+\widehat{h}_1+\widehat{h}_{4}+\widehat{h}_{5}+\widehat{h}_{6}\right]+S^\phi_1,\nonumber\\
\widetilde{\overline{g}}_{2}&=&\overline{g}_{2}+\left[\widehat{h}_0-\widehat{h}_1+\widehat{h}_{4}+\widehat{h}_{5}+\widehat{h}_{6}\right]+S^\phi_2,\nonumber\\
\widetilde{\overline{g}}_{3}&=&\overline{g}_{3}+\left[\widehat{h}_0+\widehat{h}_2-\widehat{h}_{4}+\widehat{h}_{5}+\widehat{h}_{6}\right]+S^\phi_3,\nonumber\\
\widetilde{\overline{g}}_{4}&=&\overline{g}_{4}+\left[\widehat{h}_0-\widehat{h}_2-\widehat{h}_{4}+\widehat{h}_{5}+\widehat{h}_{6}\right]+S^\phi_4,\nonumber\\
\widetilde{\overline{g}}_{5}&=&\overline{g}_{5}+\left[\widehat{h}_0+\widehat{h}_3-2\widehat{h}_{5}+\widehat{h}_{6}\right]+S^\phi_5,\nonumber\\
\widetilde{\overline{g}}_{6}&=&\overline{g}_{6}+\left[\widehat{h}_0-\widehat{h}_3-2\widehat{h}_{5}+\widehat{h}_{6}\right]+S^\phi_6.
\end{eqnarray}

\newpage
\section*{References}
%\bibliography{paper}

\end{document}